\documentclass[11pt]{JHEP3} 

\JHEPspecialurl{http://jhep.sissa.it/JOURNAL/JHEP3.tar.gz}
\usepackage{epsfig,multicol}
\usepackage{graphicx}

\usepackage{amsmath}
\usepackage{amssymb}
\usepackage{accents}

\usepackage{slashed}

\DeclareFontFamily{OT1}{rsfs}{}
\DeclareFontShape{OT1}{rsfs}{m}{n}{<-7> rsfs5 <7-10> rsfs7 <10->rsfs10}{}
\DeclareMathAlphabet{\mycal}{OT1}{rsfs}{m}{n}

\newenvironment{definition}[1][Definition]{\begin{trivlist}
\item[\hskip \labelsep {\bfseries #1}]}{\end{trivlist}}

\newcommand{\dd}{{\rm d}}

\def\dual#1{\accentset{\boldsymbol{\neg}\vspace{-0.2ex}}{#1}}
\def\opp#1{\mathsf{#1}}
\def\oppD#1{\dual{\mathsf{#1}}}
\def\oppL#1{\overleftarrow{\mathsf{#1}}}
\def\oppDL#1{\overleftarrow{\dual{\mathsf{#1}}}}
\newcommand{\be}{\begin{eqnarray}}
\newcommand{\ee}{\end{eqnarray}}
\newcommand{\ba}{\left( \begin{array}{cc}}
\newcommand{\ea} {\end{array} \right)}
\newcommand{\bv}{\left( \begin{array}{c}}
\newcommand{\ev} {\end{array} \right)}
\newcommand\fverb{\setbox\fverbbox=\hbox\bgroup\verb}
\newcommand\fverbdo{\egroup\medskip\noindent%
			\fbox{\unhbox\fverbbox}\ }
\newcommand\fverbit{\egroup\item[\fbox{\unhbox\fverbbox}]}
\newbox\fverbbox

\title{Dark spinor models in gravitation and cosmology}

\author{Christian G.~B\"ohmer\\
\email{c.boehmer@ucl.ac.uk}\\
{Department of Mathematics and Institute of Origins,
University College London, Gower Street, London, WC1E 6BT, United Kingdom}}

\author{James Burnett\\
\email{j.burnett@ucl.ac.uk}\\
{Department of Mathematics and Institute of Origins,
University College London, Gower Street, London, WC1E 6BT, United Kingdom}}

\author{David F.~Mota\\
\email{d.f.mota@astro.uio.no}\\
{Institute of Theoretical Astrophysics, University of Oslo,
0315 Oslo, Norway}}

\author{Douglas J. Shaw\\
\email{D.Shaw@qmul.ac.uk}\\
{Queen Mary University of London, Astronomy Unit,
Mile End Road, London E1 4NS, United Kingdom}}

\received{\today} 
\accepted{} 

\abstract{We introduce and carefully define an entire class of field theories based on non-standard spinors. Their dominant interaction is via the gravitational field which makes them naturally dark; we refer to them as \emph{Dark Spinors}. We provide a critical analysis of previous proposals for dark spinors noting that they violate Lorentz invariance.  As a working assumption we restrict our analysis to non-standard spinors which preserve Lorentz invariance, whilst being non-local and explicitly construct such a theory. We construct the complete energy-momentum tensor and derive its components explicitly by assuming a specific projection operator. It is natural to next consider dark spinors in a cosmological setting. We find various interesting solutions where the spinor field leads to slow roll and fast roll de Sitter solutions. We also analyse models where the spinor is coupled conformally to gravity, and consider the perturbations and stability of the spinor.}

\preprint{}

\begin{document}

\section{Introduction}
\label{sect:Introduction}

In recent years, our understanding of the universe has become greatly improved thanks to the high precision cosmological observations that we have available today. According to the Standard Model of Cosmology, which assumes General Relativity as the theory describing the gravitational interaction, our universe is composed by about $4\%$ of baryons, $23\%$ of dark matter and $73\%$ of dark energy. Moreover, in addition to these components, we need to assume an early inflationary epoch in order to explain the current state of our universe. Although this budget enables us to successfully account for the current cosmological data, it needs to assume the existence of three unknown components from a particle physics point of view, namely: dark matter, dark energy and inflaton field. Thus, we find that predictions based on General Relativity plus the Standard Model of particle physics are at odds with current astronomical observations, not only on cosmological scales, but also on galactic scales where dark matter plays a crucial role. This indicates failures either in particle physics or in general relativity (or both) and, in particular, it might be indicating the existence of new particles/fields as candidates to dark matter, dark energy and the inflaton which could arise in high energy physics \cite{Li4,Li5,Bi1,Li1,Li2,Ci1,Ci2,Ci3,Li3,Bi2,Ci4,Ci5,Bi3,Watanabe:2009nc,Poplawski2010,Mota:2007sz,alpha}.

Spinors have played an important role in mathematics and physics throughout the last 80 years. They theoretically model particles with half integer spin, like the electron in the massive case or the neutrino (massive or massless). The spin structure of manifolds has played an important part in modern mathematics, while in mathematical physics this structure motivated the twistor program.

In the framework of particle physics all spinors used are either Dirac, Weyl (massless Dirac spinors) or Majorana spinors, $\psi$. Such spinors obey a field equation which is first order in the derivatives (momenta) of $\psi$. Cosmologically, this first order field equation implies that the average value of both $\Phi = \bar{\psi}\psi$ and the spinor energy density of a free spinor field evolves like the energy density of pressure-less dust i.e.~proportional to $(1+z)^{3}$, where $z$ is the redshift. Additionally, the first order nature of the field equation results in a quantum propagator, $G_{F}$, which, for large momenta $p$, behaves as $G_{F} \propto p^{-1}$. This limits the form of perturbatively renormalizable spinor self-interaction terms in the action to be no more than quadratic in $\psi$ e.g.~$\bar{\psi}\psi$ and $\bar{\psi}\gamma_{\mu}A^{\mu}\psi$. The momentum drop-off of $G_{F}$ also results in $\psi$ having a canonical mass dimension of $3/2$. 

A wider range of renormalizable self-interaction terms and cosmological behavior would be allowed if one could construct a viable spinor field theory where $G_F \propto p^{-2}$, for large $p$, resulting in a $\psi$ with a canonical mass dimension of unity. We refer to this entire class of spinor field theories with such properties as Non-Standard Spinors (NSS). This class of spinors is closely related to Wigner's non-standard classes~\cite{Wigner:1939cj}. Weinberg showed that, under the assumptions of Lorentz invariance and locality, the only spin-$1/2$ quantum field theory is that which describes standard spinors (Dirac, Weyl, Majorana). NSS will therefore violate either locality or Lorentz invariance, or possibly both. Our working assumption is that reasonable NSS models preserve Lorentz invariance, while being non-local.

Along these lines of reasoning, Ahluwalia-Khalilova and Grumiller~\cite{jcap,prd} constructed a NSS model using momentum space eigen-spinors of the charge conjugation operator \emph{Eigenspinoren des LadungsKonjugationsOperators} (ELKO) to build a quantum field. They showed that such spinors belong to a non-standard Wigner class, and to exhibit non-locality~\cite{Wigner:1939cj}. They satisfy $(CPT)^2=-\mathbb{I}$ while Dirac spinors satisfy $(CPT)^2=\mathbb{I}$. In more mathematical terms, they belong to a wider class of spinorial fields, so-called flagpole spinor fields~\cite{daRocha:2005ti}. The spinors correspond to the class 5, according to Lounesto's classification which is based on bilinear covariants, similar to Majarona spinors, see also~\cite{daRocha:2008we,daRocha:2009gb,HoffdaSilva:2009is}. Locality issues and Lorentz invariance were further investigated in~\cite{Ahluwalia:2008xi,Ahluwalia:2009rh} resulting in results along the lines of the current work. Causality has been analyzed in~\cite{Fabbri:2009ka,Fabbri:2009aj}.

The construction of ELKOs using momentum space eigenspinors, $\lambda(\mathbf{p},h,e)$, of the charge conjugation operator leads to a spinor field with a double helicity structure. The left-handed and the right-handed spinor have opposite helicities which in turn requires a careful construction of the resulting field theory. These spinors have received quite some attention recently~\cite{Boehmer:2006qq,Boehmer:2007dh,n6} and their effects in cosmology have been investigated~\cite{Boehmer:2007ut,Boehmer:2008ah,Boehmer:2008rz,Gredat:2008qf,Boehmer:2009aw,n1,n3,n4,n5,Shankaranarayanan:2009sz,Shankaranarayanan:2010st,Boehmer:2010tv,Wei:2010ad}.

However, as we will show in \S \ref{sec:sNSS}, ELKOs spinors, defined in the way described above, are not Lorentz invariant. We demonstrate using our construction of NSS where this Lorentz violation appears, thus confirming~\cite{Ahluwalia:2008xi,Ahluwalia:2009rh}. The original analyses defined the field structure entirely in terms of momentum space basis spinors rather than say starting with an action whose minimization would imply that structure. This led to the violation of Lorentz invariance being hidden in the mathematical structure of the model. In the present work, on the other hand, we start with a general action principle for NSS. When applied to the  ELKOs and an alternative model also based eigenspinors of the charge conjugation operator, the violation of Lorentz invariance and other issues with their construction are explicit at the level of the action. The original ELKO definition is seen to require a preferred space-like direction and is ill-defined when the momentum points along that direction. We offer a new NSS field theory  which is also based on the eigenspinors of the charge conjugation operator (i.e.~using the basis $\lambda(\mathbf{p},h,e)$) which respects the rotational group $SO(3)$ but is not invariant under boosts.

We shall see that the general construction of NSS models can be seen as the choice of some operator $\opp{P}$ satisfying $\opp{P}^2 = \mathbb{I}$ which acts on $\psi$ to project out what states that would otherwise give an inconsistent Hamiltonian density. In this article we provide a general treatment of class of NSS models based on an action principle and choice of operator $\opp{P}$. We show that there is one, potentially unique, choice of $P$ which results in a Lorentz invariant, ghost-free but non-local spinor field theory with canonical mass dimension one. 

We are also interested in the cosmological behavior of general NSS models and construct the energy momentum tensor, $T_{\mu\nu}$. For ELKO spinors it appears that, at present, no one has obtained the full $T_{\mu\nu}$ as all previous works in the literature, including ours, have overlooked contributions to $T_{\mu\nu}$ from the variation of spin connection.

This article is organized as follows: we define our notation, general spinors and exactly what a non-standard spinor is in \S \ref{sec:spinors}, then in \S\ref{sec:sNSS} we look specifically at the original ELKO definition, offer a modified version, finishing the section by examining the possibility of a Lorentz invariant non-standard spinor. In \S \ref{sec:emt} we examine the energy momentum tensor both with and without the projection operator, this then leads us nicely into sections \S \ref{sec:ELKOCos} and \S \ref{sec:NSSCos}, where we examine the cosmological applications of both the original ELKO and the Lorentz invariant NSS respectively and in each case note the existence of non-trivial de Sitter solutions. We make our final remarks in \S \ref{sec:conc}, followed by three appendices showing explicit calculations of the variation of the spin connection with respect to the metric for the general case, the Dirac spinor and finally the ELKO spinor.

\section{Generalized Spinor Actions}
\label{sec:spinors}

\subsection{Notation and Preliminaries}

We work with a metric signature $(+,-,-,-)$, and define $\gamma$-matrices, $\gamma^{a}$, in the Weyl basis:
\be
\gamma^{0} = \begin{pmatrix} 0 & \mathbb{I}_{2\times 2} \\ \mathbb{I}_{2\times 2} & 0 \end{pmatrix}, \qquad
\gamma^{i} = \begin{pmatrix} 0 & -\sigma^{i} \\
 \sigma^{i} & 0 \end{pmatrix},
\ee
where $\sigma^{i}$ are the Pauli matrices:
\be
\sigma^{1} = \begin{pmatrix} 0 & 1 \\ 1 & 0 \end{pmatrix}, \qquad
\sigma^{2} = \begin{pmatrix} 0 & -i \\ i & 0 \end{pmatrix}, \qquad
\sigma^{3} = \begin{pmatrix} 1 & 0 \\ 0 & -1 \end{pmatrix}.
\ee
We also define tetrads $e_{\mu}^{a}$ by $e_{\mu}^{a}e_{\nu}^{b}\eta_{ab} = g_{\mu \nu}$, where $g_{\mu \nu}$ is the space-time metric and $\eta_{ab} = {\rm diag}(1,-1,-1,-1)$. Space-time $\gamma$-matrices, $\gamma^{\mu}$ are then given by $\gamma^{\mu} = e^{\mu}_{a}\gamma^{a}$, and hence obey:
\be
\gamma^{\mu}\gamma^{\nu} +\gamma^{\nu}\gamma^{\mu} = 2g^{\mu \nu}. \nonumber
\ee
We also define $\gamma^{5} = i \gamma^{0}\gamma^{1}\gamma^{2}\gamma^{3}$ i.e.:
\be
\gamma^{5} = \begin{pmatrix} \mathbb{I}_{2\times 2} & 0 \\ 0 & -\mathbb{I}_{2\times 2} \end{pmatrix}.
\ee
The covariant derivative $\nabla_{\mu}$ is defined by $\nabla_{\mu}g_{\nu \rho} = 0$, and so acting on a vector $A^{\mu}$ one has:
\be
\nabla_{\mu} A^{\nu} = \partial_{\mu} A^{\nu} + \Gamma_{\mu \rho}^{\nu} A^{\rho},
\ee
where $\Gamma_{\mu \rho}^{\nu}$ denotes the Christoffel symbol of $g_{\mu \nu}$. The definition of $\nabla_{\mu}$ is extended to spinors by further requiring that $\nabla_{\mu} e_{\nu}^{a} = 0$; hence $\nabla_{\mu}\gamma^{\nu} = 0$. The extension defines the spin connection:
\be
\omega_{\mu}^{ab} = e_{\nu}^{a}\partial_{\mu}e^{\nu b} + e_{\nu}^{a}e^{\sigma b} \Gamma_{\mu \sigma}^{\nu}.
\ee
The action of $\nabla_{\mu}$ on a spinor $\psi$ is then given by:
\be
\nabla_{\mu} \psi \equiv \partial_{\mu} \psi - \Gamma_{\mu} \psi
\ee
where $\Gamma_\mu$ is given by:
\be
\label{eq:spincon}
\Gamma_{\mu} = \frac{i}{4}\omega^{ab}_{\mu} f_{ab}, \qquad
f^{ab} = \frac{i}{2}\left[ \gamma^{a}, \gamma^{b}\right].
\ee
The adjoint of an arbitrary spinor is defined by $\bar{\psi} = \psi^{\dagger}\gamma^{0}$, where $\psi^{\dagger}$ is the hermitian conjugate of $\psi$. Since $\bar{\psi}\psi$ is a space-time scaler, it follows that $\nabla_{\mu}$ acts on adjoint spinors thus:
\be
\bar{\psi} \overleftarrow{\nabla_{\mu}} \equiv \nabla_{\mu} \bar{\psi} \equiv \partial_{\mu} \bar{\psi} + \bar{\psi} \Gamma_{\mu}.
\ee
Similarly, if we define a dual spinor $\dual{\psi}$ of $\psi$ so that $\dual{\psi}\psi$ is a space-time scalar, we have:
\be
\dual{\psi} \overleftarrow{\nabla}_{\mu} \equiv \nabla_{\mu} \dual{\psi} \equiv \partial_{\mu} \dual{\psi} + \dual{\psi} \Gamma_{\mu}.
\ee
We also define the slashed notation thus: $\slashed{A} = \gamma^{\mu}A_{\mu}$ so $\slashed{\nabla} = \gamma^{\mu}\nabla_{\mu}$. The usual Dirac dual spinor is $\bar{\psi} = \psi^{\dagger}\gamma^{0}$. 

We for any operator $\opp{A}$, acting on the right or $\oppL{A}$, acting on the left, we define the respective dual operators $\oppDL{A}$ and $\oppD{A}$  by the requirement that $\dual{\psi} \oppDL{A}$ be dual to $\opp{A} \psi$  and that  $\oppD{A} \psi$ be dual to $\dual{\psi} \oppL{A}$ for any $\psi$. We note that $\oppDL{A}$ acts on the left and $\oppD{A}$ acts on the right.

The commutator of two covariant derivatives on a spinor can then be calculated to be:
\be
[\nabla_{\mu},\nabla_{\nu}]\psi &=& \left[\partial_{\nu}\Gamma_{\mu}-\partial_{\mu}\Gamma_{\nu}+\Gamma_{\mu}\Gamma_{\nu}-\Gamma_{\nu}\Gamma_{\mu}\right]\psi \nonumber \\ &=& \frac{1}{8} R_{\mu \nu \rho \sigma} \left[ \gamma^{\rho}, \gamma^{\sigma}\right] \psi, = -\frac{i}{4} R_{\mu \nu \rho \sigma} f^{\rho \sigma} \psi,
\ee
where $R_{\mu \nu\rho\sigma}$ is the Riemann curvature tensor, see e.g.~\cite{Hehl:1976kj}. It follows that:
\be
\slashed{\nabla}^2 \psi = \nabla^2 \psi + \mathcal{R}\psi.
\ee
where
\be
\mathcal{R} = -\frac{1}{4}R_{\mu\nu\rho\sigma}f^{\mu \nu}f^{\rho \sigma} = \frac{1}{4}R_{\mu \nu \rho \sigma}\gamma^{\mu}\gamma^{\nu}\gamma^{\rho}\gamma^{\sigma}.
\ee

\subsection{Generalized Free Spinor Actions}

We begin with the criterion that a free, massive spinor free field, $\psi$, in flat space-time (with tetrads $e_{\mu}^{a} = \delta_{\mu}^{a}$ so $\Gamma_{\mu} = 0$) should obey the flat space Klein-Gordon equation:
\be
\label{flatKleinG}
\partial^2 \psi = m^2_{\psi} \psi.
\ee
This suggests the following flat-space Lagrangian for $\psi$:
\be
\mathcal{L}_{\rm free-flat}^{(1)} &\equiv& ( \dual{\psi}\overleftarrow{\slashed{\partial}})(\slashed{\partial}\psi) - m^2_{\psi}\dual{\psi}\psi,
\ee
where $\dual{\psi}$ is some dual spinor to $\psi$ defined so that $\dual{\psi}\psi$ is a space-time scalar. We vary $\psi$ and $\dual{\psi}$ independently. We note that up to a surface term, the above action, $\mathcal{L}_{\rm free-flat}^{(1)}$ is equivalent to another $\mathcal{L}_{\rm free-flat}^{(2)}$ given by:
\be
\mathcal{L}_{\rm free-flat}^{(2)} &\equiv& (\partial_{\mu} \dual{\psi})(\partial^{\mu}\psi) - m^2_{\psi}\dual{\psi}\psi.
\ee
However, this equivalence relies on $\partial^2\psi = \slashed{\partial}^2\psi$ which is broken when the actions are promoted to curved space by taking $\partial_{\mu} \rightarrow \nabla_{\mu}$, since generally $\mathcal{R} \neq 0$ when $R_{\mu \nu \rho \sigma} \neq 0$. One must therefore choose which of the two actions to promote to curved space.

Remaining in flat-space, there is a problem with both actions as there are given above. The field equation $(\partial^2 -m^2)\psi = 0$ constrains the evolution of each of the four components of $\psi$ but does not impose any relation between the different components. We define a basis $\psi_{i}$ where $i = 1,2,3,4$ on 4-spinor space, such that, $\dual{\psi}_{i}\psi_{j} = 0$ if $i \neq j$ and $\partial_{\mu}\psi_{i} = 0$. We assume that $\partial_{\mu}\dual{\psi}_{i} = 0$. However, as is well known, Lorentz invariance prevents us from defining $\dual{\psi}_{i}\psi_{j} = \delta_{ij}$, instead we can ensure that $\dual{\psi}_{1}\psi_{1} = \dual{\psi}_{2}\psi_{2} = 1$ and $\dual{\psi}_{3}\psi_{3} = \dual{\psi}_{4}\psi_{4} = -1$. Solutions of $(\partial^2 -m^2)\psi = 0$ are then given by:
\be
\psi = \sum_{i, \mathbf{p}} a_{i}(\mathbf{p}) \frac{1}{2E_{p}} e^{i E_{p} t - i\mathbf{p}\cdot \mathbf{x}} \psi_{i} + \sum_{i, \mathbf{p}} b^{\dagger}_{i}(\mathbf{p}) \frac{1}{2E_{p}} e^{-i E_p t + i\mathbf{p}\cdot \mathbf{x}} \psi_{i},\nonumber
\ee
where $a_{i}(\mathbf{p})$ and $b^{\dagger}_{i}(\mathbf{p})$ are some functions of $\mathbf{p}$ and $E_{p} = \sqrt{m^2+\mathbf{p}^2}$. Here $\sum_{p} = \int \dd^{3} p$.

Let us define the Hamiltonian density $\mathcal{H} = \dot{\dual{\psi}}\dual{\pi} + \pi\dot\psi - \mathcal{L}^{(1)}$
where the momentum is defined as usual $\pi = \partial \mathcal{L}^{(1)}/ \partial \dot{\psi} = \dot{\dual{\psi}}$, and $\dual{\pi} = \partial \mathcal{L}^{(1)}/ \partial \dot{\dual{\psi}} = \dot{\psi}$.  In flat-space, the Hamiltonian density formed from on $\mathcal{L}^{(2)}$ differs from that based on $\mathcal{L}^{(1)}$ only by an irrelevant total derivative which can be dropped. We then have
\be
\mathcal{H}=\left[\pi \dual{\pi} + \nabla^i\dual{\psi}\nabla_i\psi + m^2 \dual{\psi}\psi\right].
\ee
Taking $\epsilon_{i} = \dual{\psi}_{i}\psi_{i}$, one can show that
\be
H = \int \dd^3 x\, \mathcal{H} = \sum_{j} \epsilon_{j}\sum_{\mathbf{p}}\frac{(E_{p}^2 +\mathbf{p}^2+m^2)}{2E_{p}} [a^{\dagger}_{j}(\mathbf{p})a_{j}(\mathbf{p}) + b_{j}(\mathbf{p})b^{\dagger}_{j}(\mathbf{p})]
\ee
which then becomes
\be
H = \sum_{j}\epsilon_{j}\sum_{\mathbf{p}}(E_{p}) [a^{\dagger}_{j}(\mathbf{p})a_{j}(\mathbf{p}) + b_{j}(\mathbf{p})b^{\dagger}_{j}(\mathbf{p}) ].
\ee

Finally we can assume that these will be upgraded to operators and since we are referring to spin one half particles we are dealing with fermions and therefore anti-commutation.
\be
H = \sum_{j}\epsilon_{j}\sum_{\mathbf{p}}(E_{p}) [a^{\dagger}_{j}( \mathbf{p})a_{j}(\mathbf{p}) - b^{\dagger}_{j}(\mathbf{p})b_{j}(\mathbf{p}) ].
\ee
This then gives an ill defined Hamiltonian density which is not positive definitive. However, we know that if we were to write the Dirac spinor in the KG equation and followed the same step we would get a consistent Hamiltonian density. Thus, there is a projection operation implicitly present which removes (projects out) the components of the spinor which would give an inconsistent Hamiltonian density. It is important to note that this not directly related to the actual energy as the energy is squared in this expression and therefore we retain the negative energy information, which is, of course, what we learned from Dirac.

Let us assume that the $a_{i}$ and $a_{i}^{\dagger}$ to represent creation and annihilation operators, then $a_{i}^{\dagger}a_{i} \neq 0 $ and $b_{i}^{\dagger}b_{i} \neq 0$. If we interpret $\dual{\psi}\psi$ as the energy-density of the spinor field with $\epsilon_{1}=\epsilon_{2} = -\epsilon_{3}=-\epsilon_{4} = 1$, it follows that the spinor field can have negative energy density, unless there is some additional condition that requires $a_{3} = a_{4} = 0$ and $b_{1}=b_{2}=0$ in the definition of $\psi$. Additionally, without such a requirement it would be possible to have states with both $a_{i}^{\dagger}a_{i}$ and $b_{i}^{\dagger}b_{i} \geq 0$ but with zero energy. Negative energy or ghost states lead to well known instabilities both classically and at the level of quantum field theory.

The requirement that $a_{3}=a_{4}=0$ and $b_{1}=b_{2}=0$ can be seen as an additional equation for $\psi$ which projects out negative energy states, i.e. we would have $\opp{P} \psi = \psi$ for some operator $\opp{P}$ with the property:
\be
\opp{P} \left(\psi_{i} e^{\mp i p_{\mu}x^{\mu}}\right) = \mp \epsilon_{i} \psi_{i} e^{\mp i p_{\mu}x^{\mu}}.
\label{projection}
\ee
where $p^{\mu} = (E_{p},\mathbf{p}^{i})$. This form implies that when one moves to momentum space $P(p^{\mu})$ is an odd function of $p^{\mu}$. If one were to attempt to define spinors using a $\opp{P}(p^{\mu})$ that was an even function of $p^{\mu}$, one would have to require that the $a_{i}(\mathbf{p})$ and $b_{i}(\mathbf{p})$ commute rather than anti-commute leading to a field obeying Bose-Einstein statistics.

We define projection operators:
\be
\opp{P}_{\pm} = \frac{1}{2}\left[\mathbb{I} \pm \opp{P}\right],
\ee
and note that:
\be
\opp{P}_{\pm}\opp{P}_{\pm} \equiv \opp{P}_{\pm}, \quad \opp{P}_{\pm}\opp{P}_{\mp} \equiv 0.
\ee
Positive / negative energy spinor modes then respectively correspond to those which obey $\opp{P}_{+}\psi = \psi$ and $\opp{P}_{-}\psi = \psi$, or written out explicitly
\be
\opp{P}_+(\psi) = \frac12 \sum_{i, \mathbf{p}}(1+\epsilon_i) a_{i}(\mathbf{p}) e^{i E_{p} t - i \mathbf{p}\cdot \mathbf{x}} \psi_{i} +\frac12 \sum_{i, \mathbf{p}} (1-\epsilon_j) b^{\dagger}_{i}(\mathbf{p}) e^{-i E_p t + i\mathbf{p} \cdot \mathbf{x}} \psi_{i}.
\ee
We also define the adjoint operator, $\oppL{P}$, by:
\be
\int \sqrt{-g}{\rm d}^4 x\, \dual{\chi} \opp{P} \psi \cong \int \sqrt{-g}{\rm d}^4 x\, \dual{\chi} \oppL{P} \psi ,
\ee
where $\cong$ implies that this relation is true up to a surface integral term.   We let $\oppL{P}_{\pm} = (\mathbb{I}\pm \oppL{P})/2$. For any $\psi$ and $\dual{\psi}$ we then define the shorthand:
\be
\psi_{\pm} = \opp{P}_{\pm}\psi, \quad \dual{\psi}_{\pm} = \dual{\psi} \oppL{P}_{\pm}.
\ee
We can now rewrite our Hamiltonian density as
\be
H = \frac14 \sum_{j, \mathbf{p}}E_{p} \left[\epsilon_{j}(1+\epsilon_j)^2a^{\dagger}_{j}(\mathbf{p})a_{j}(\mathbf{p}) - \epsilon_{j}(1-\epsilon_j)^2b^{\dagger}_{j}(\mathbf{p})b_{j}(\mathbf{p}) \right].
\ee
One sees that if we use the definition laid out earlier, namely $\epsilon_1 =\epsilon_2=-\epsilon_3=-\epsilon_4=1 $, we find that this Hamiltonian density becomes
\be
H =  \sum_{j=1, \mathbf{p}}^2 E_{p} a^{\dagger}_{j}(\mathbf{p})a_{j}(\mathbf{p}) + \sum_{j=3, \mathbf{p}}^4 E_{p} b^{\dagger}_{j}(\mathbf{p})b_{j}(\mathbf{p}),
\ee
which is now positive definite for any spinor field, provided it satisfies the projection condition~(\ref{projection}).

Next we discuss and alternative approach to enforce the condition that only positive energy modes propagate i.e.~$\psi = \psi_{+}$ and $\dual{\psi} = \dual{\psi}_{+}$, which is equivalent to $\opp{P}_{-}\psi = 0$ and $\dual{\psi} \oppL{P}_{-} = 0$. Suppose that we initially take the Lagrangian density for $\psi$ and $\dual{\psi}$ to be $\mathcal{L}_{\psi}(\psi, \dual{\psi})$. We may project out unphysical modes by adding an extra term, $\mathcal{L}_{P}$, to the Lagrangian: $\mathcal{L}_{\psi} \rightarrow \mathcal{L}_{\psi} + \mathcal{L}_{P}$ where:
\be
\mathcal{L}_{P} &=& -\dual{\chi} \opp{P}_{-} \psi - \dual{\psi} \oppL{P}_{-}\chi, 
\\ &\cong&  -\dual{\chi} \oppL{P}_{-} \psi - \dual{\psi} \opp{P}_{-}\chi, \nonumber
\ee
where $\cong$ indicates equality up to a total derivative.  

Varying the action with respect to $\chi$ then gives $\psi_{-} = \dual{\psi}_{-} = 0$, as required. With $\mathcal{S}_{\psi} = \int {\rm d}^4 x \sqrt{-g} \mathcal{L}_{\psi}$, the other field equations from the variation of the action with respect to $\psi$ and $\dual{\psi}$ are:
\be
\left[\frac{\delta \mathcal{S}_{\psi}}{\delta \dual{\psi}}\right]_{+} = 0, \quad
\left[\frac{\delta \mathcal{S}_{\psi}}{\delta \psi}\right]_{+} = 0, \\
\left[\frac{\delta \mathcal{S}_{\psi}}{\delta \dual{\psi}}\right]_{-} = \chi_{-}, \quad
\left[\frac{\delta \mathcal{S}_{\psi}}{\delta \psi}\right]_{-} = \dual{\chi}_{-},
\ee
and $\chi_{+}$, $\dual{\chi}_{+}$ are undetermined gauge degrees of freedom which do not enter the action or field equations. We may integrate out the $\chi$ fields by replacing $\psi$ and $\dual{\psi}$ with $\psi_{+}$ and $\dual{\psi}_{+}$ in the Lagrangian $\mathcal{L}_{\psi}$, i.e. we redefine:
\be
\mathcal{L}_{\psi}(\psi, \dual{\psi}) \rightarrow \mathcal{L}_{\psi}(\psi_{+},\dual{\psi}_{+}).
\ee
Then the field equations follow from:
\be
\left[\frac{\delta \mathcal{S}_{\psi}}{\delta \dual{\psi}_{+}}\right]_{+} = \left[\frac{\delta \mathcal{S}_{\psi}}{\delta \psi_{+}}\right]_{+} = 0.
\ee

Now we know that:
\be
\opp{P}^2 \left(\psi_{i} e^{\mp i p_{\mu}x^{\mu}}\right) = \epsilon_{i}^2 \psi_{i} e^{\mp i p_{\mu}x^{\mu}} = \psi_{i} e^{\mp i p_{\mu}x^{\mu}},
\ee
and so we must have that $\opp{P}^2 \cong \mathbb{I}$ where $\cong$ implies that this identity holds modulo the field equation $p^2 = m^2$. We also know that $\opp{P}^2$ is an even function of $p$ and $\opp{P}(p)$ is an odd function, so we may write $\opp{P}^2 = \opp{F}(p^2/m^2-1)$ where $\opp{F}$ is some operator which depends on $p^2/m^2$; we must then have $\opp{F}(0) = \mathbb{I}$. If $\opp{F}(p^2/m^2-1)= \mathbb{I}$ implies that $p^2/m^2 = 1$, e.g. if $\opp{P}^2 = \opp{F} = p^2/m^2$, then $\opp{P}\psi = \psi$ implies the field equation $p^2 \psi = m^2 \psi$ rendering the latter superfluous. It would then be sufficient to take the total action to be simply $\mathcal{L}_{P}$:
\be
\mathcal{L}_{\psi} = \mathcal{L}_{P}.
\ee
This is precisely what happens for Dirac and Majorana fields where, respectively $\opp{P}$ is $\opp{P}_{D} = i \slashed{\nabla}/m$ and $\opp{P}_{M} = \opp{C} i\slashed{\nabla}/m$ and $\opp{C}\psi = \psi^{c}$; here $\psi^{c}$ is the charge conjugate spinor field and $\opp{C}^2 = \mathbb{I}$. Positivity of the energy then requires $\dual{\chi}=m\bar{\psi} = m\psi^{\dagger}\gamma^{0}$ in the Dirac case, and $\dual{\chi} = m\bar{\psi} \opp{C}$ in the Majorana case.

\subsection{Non-Standard Spinors}

We shall think of Dirac and Majorana as standard classes of spinors. In flat space, we define a general class free non-standard spinors (NSS) as being those spinor which, in momentum space, obey:
\be
p^2 \psi(p) &=& m^2 \psi(p), \label{psiEqn1} \\
\opp{P}(p^{\mu})\psi(p) &=& \psi(p),\label{psiEqn2}
\ee
where $\opp{P}(p^{\mu}) = -\opp{P}(-p^{\mu})$ and where $\opp{P}^2(p^{\mu}) = \mathbb{I}$ does \emph{not} automatically imply $p^2 = m^2$, so that Eqs.~\eqref{psiEqn1} and \eqref{psiEqn2} are independent, the former fixing the dynamics of $\psi$ and the latter the spinor structure. One such form for the $\opp{P}$ operator would be:
\be
\opp{P}(p) = \sin\left( \frac{\pi \slashed{p}}{2 m}\right), \nonumber
\ee
thus $\opp{P}^2 = \sin^2(\pi \sqrt{ p_{\mu}p^{\mu}} / 2 m)$ and $\opp{P}^2=\mathbb{I}$ only implies $\sqrt{p_{\mu}p^{\mu}} / 2m = 2n+1$ for $n \in \mathbb{N}$. However, whilst this does not imply $p^2 = m^2$ globally, it does require $p^2=m^2$ locally. This is to say that in momentum space, for $p^{\mu}$ lying in or close to the sub-space, $S_{p^2=m^2}$, of points defined by $p^2 = m^2$, $P^2 = \mathbb{I}$ requires that $p^{\mu}$ be in $S_{p^2=m^2}$.  So close to $S_{p^2=m^2}$, the Eq.~\eqref{psiEqn1} is again superfluous. We therefore further require, in our definition of non-standard spinors, that in some open region around the sub-space $S_{p^2=m^2}$, we have $\opp{P}^2(p) \equiv \mathbb{I}$. For simplicity we may therefore restrict to consider $\opp{P}(p)$ such that $\opp{P}^2(p) = \mathbb{I}$ for all $p^{\mu}$. 

We would also like non-standard spinors to have a canonical mass dimension of unity, like scalar fields rather than the $3/2$ mass dimension of Dirac / Majorana spinors. The canonical mass dimension of a quantum field is determined by the momentum drop-off of the free field propagator, $G_{F}(p;m)$, for $\vert p^2 \vert \gg m^2$. For standard spinors $G_{F}(p;m) \sim O(p^{-1})$ whereas for scalar fields or vector bosons, both with mass dimension one, $G_{F}(p;m) \sim O(p^{-2})$. In general, if $G_{F}(p;m) \sim O(p^{-2+\delta})$ the quantum field has canonical mass dimension $1+\delta/2$. This definition of the mass dimensions also determines the renormalizability of self-interaction terms. 
For a general field $\Psi$ (not necessarily a spinor), by counting powers of momentum in field loops, one determined that if $G_{F}(p;m) \sim O(p^{-2+\delta})$ the self-interaction terms of $O(\Psi)^{n}$ are \emph{not} perturbatively renormalizable in $3+1$ dimensions if $n > 4/(1+\delta/2)$. With spinor fields self interactions must all involve an equal number of $\psi$ and $\dual{\psi}$ fields and so $n$ must be even. Thus if the mass dimensions is $3/2$ then we could only have $n \leq 2$ (as $n=3$ is not allowed) implying that only renormalizable self-interaction terms are simply mass terms proportional to $\dual{\psi}\psi$. However a mass dimensions one NSS field ($\delta =0$) could be renormalizable with a fourth order interaction term ($n=4$) and so we could have additional self-interaction terms of the form $\lambda(\bar{\psi}\psi)^2$.  We shall see that $\delta =0$ requires $\lim_{\vert p^2 \vert \gg m^2} \opp{P}(p^{\mu}) \sim O(\vert p^2\vert^{-n/2})$ or equivalently $\lim_{\lambda \rightarrow \infty}\opp{P}(\lambda p^{\mu}) \sim O(\lambda^{-n})$ for some $n \leq 0$.  We note that this condition (with $n=0$) is implied by the requirement that $\opp{P}^2(p^{\mu}) = \mathbb{I}$ for all $p^{\mu}$.

Finally $\opp{P}$ must be chosen so that the NSS spinor action is real (or at least real up to a surface integral).  Firstly this implies that the dual spinor must be defined so that $(\dual{\psi}\psi)^{\dagger} = \dual{\psi}\psi$ for any $\psi$.  Reality of the kinetic term in the action requires that either $\dual{\slashed{\nabla}}^2 = \overleftarrow{\slashed{\nabla}}^2$ or $\dual{\nabla}^2 = \overleftarrow{\nabla}^2$, depending on the choice of kinetic structure.  Finally reality of the projection term $\mathcal{L}_{P}$ requires that:
\be
\opp{P} = \oppD{P}. \nonumber
\ee

We summarize the definition of NSS below.

\begin{definition}
A non-standard spinor, $\psi$, is defined by a operator $\opp{P}(x)$, which in momentum space is $\opp{P}(p^{\mu})$, and has the following properties:
\begin{enumerate}
\item $\opp{P}(p^{\mu})$ is an odd function of momentum: $P(p^{\mu}) = P(-p^{\mu})$. 
\item $\opp{P}^2 \equiv \mathbb{I}$ on any spinor (i.e. not just those that satisfy the field equation).
\item $\opp{P} = \oppD{P}$ on any spinor to ensure reality of the action.
\end{enumerate}
\end{definition}
The second condition implies that $\lambda \rightarrow \infty$, $\opp{P}(\lambda p^{\mu}) \sim O(\lambda^{0})$. The adjoint operator to $\opp{P}$ is $\oppL{P}$, and we define $\opp{P}_{\pm} = (\mathbb{I} \pm \opp{P})/2$, $\oppL{P}_{\pm} = (\mathbb{I} \pm \oppL{P})$, and $\psi_{\pm} = \opp{P}_{\pm}\psi$, $\dual{\psi}_{\pm} = \dual{\psi} \oppL{P}_{\pm}$. Physical modes are those for which $\psi = \psi_{+}$, $\dual{\psi} = \dual{\psi}_{+}$. Starting with some Lagrangian $\mathcal{L}(\psi,\dual{\psi})$ we can project out the unphysical modes either by adding:
\be
\mathcal{L}_{P} = -\dual{\chi} \opp{P}_{-} \psi - \dual{\psi} \oppL{P}_{-}\chi,
\ee
or by replacing $\mathcal{L}(\psi,\dual{\psi})$ with $\mathcal{L}(\psi_{+},\dual{\psi}_{+})$, both methods result in equivalent field equations, and, since $\mathcal{L}_{P}$ vanishes on-shell, in equivalent values of the action.

A free, non-standard spinor satisfies $\psi = \psi_{+}$ i.e. $\psi = P \psi$ and:
\be
\left[(p^2-m^2)\psi\right]_{+} = 0.
\ee
In flat-space, this NSS field equation results from two simple actions which are inequivalent in curved spacetimes. With the total Lagrangian taken to be $\mathcal{L}^{(i)}_{\psi-P} = \mathcal{L}_{\rm free}^{(i)}(\psi,\dual{\psi}) + \mathcal{L}_{P} \cong \mathcal{L}_{\rm free}^{(i)}(\psi_{+},\dual{\psi}_{+})$, the two choices for $\mathcal{L}_{\rm free}^{(i)}(\psi,\dual{\psi}) $ are:
\be
\mathcal{L}_{\rm free}^{(1)} &=& (\dual{\psi}\overleftarrow{\slashed{\nabla}})\slashed{\nabla}\psi - m^2\dual{\psi} \psi, \\
\mathcal{L}_{\rm free}^{(2)} &=& (\dual{\psi}\overleftarrow{\nabla}^{\mu} \nabla_{\mu}\psi) - m^2\dual{\psi} \psi.
\ee
The minimizing $\mathcal{L}^{(1)}_{\psi-P}$ gives:
\be
\left[\slashed{\nabla}^2 \psi_{+} +m^2 \psi_{+}\right]_{+} &=& 0, \nonumber \\
\left[\dual{\psi}_{+}\overleftarrow{\slashed{\nabla}}^2 +m^2 \dual{\psi}_{+}\right]_{+} &=& 0. \nonumber
\ee
The minimizing $\mathcal{L}^{(2)}_{\psi-P}$ gives:
\be
\left[\nabla^2 \psi_{+} +m^2 \psi_{+}\right]_{+} &=& 0, \nonumber \\
\left[\dual{\psi}_{+}\overleftarrow{\nabla}^2 +m^2 \dual{\psi}_{+}\right]_{+} &=& 0. \nonumber
\ee
In general, as we noted above, $\nabla^2 \neq \slashed{\nabla}^2$. In Appendix \ref{app:quant:free} we find that the free field  quantum propagator for $\psi$ is (in flat space) for action $S_{\psi}^{(1)}$  given by:
\be
G_{F}(p^{\mu}) = \frac{1}{2}\frac{\left(\mathbb{I} \pm \opp{P}(p)\right)}{p^2-m^2}.
\ee
It is straight-forward to check that this is also the flat-space free-field propagator for  for action $S_{\psi}^{(2)}$.

It is then clear that $G_{F} \sim p^{-2}$ for large $\vert p^2 \vert$ is equivalent to $\lim_{\lambda \rightarrow \infty}P(\lambda p^{\mu}) \sim O(\lambda^{-n})$ for some $n\geq 0$. For NSS spinors this is ensured (with $n=0$) by $P^2(p^{\mu}) = \mathbb{I}$ for all $p^{\mu}$. We note that if we took the NSS action with $P(p^{\mu}) = i\slashed{\nabla}/m$ then upon integrating out the $\chi$ and $\dual{\chi}$ we would recover the Dirac spinor action.

We can generalize the free-field actions to include self-interaction terms by replacing $m \dual{\psi}\psi$ with $V(\dual{\psi}\psi)$, so $\mathcal{L}_{\psi-P}^{(i)} = \mathcal{L}_{\psi}^{(i)}(\psi,\dual{\psi}) + \mathcal{L}_{P} \cong \mathcal{L}_{\psi}^{(i)}(\psi_{+},\dual{\psi}_{+})$:
\be
\mathcal{L}^{(1)}_{\psi}(\psi,\dual{\psi}) = (\dual{\psi}\overleftarrow{\slashed{\nabla}})\slashed{\nabla}\psi - V\left(\dual{\psi} \psi\right), \label{action1} \\
\mathcal{L}^{(2)}_{\psi}(\psi,\dual{\psi}) = (\nabla_{\mu}\dual{\psi})\nabla^{\mu}\psi - V\left(\dual{\psi} \psi\right). \label{action2}
\ee
By power-counting arguments we noted that perturbatively renormalizable $V(\dual{\psi}\psi)$ will have the form $$V(\dual{\psi}\psi) = V_{0} + m^2 \dual{\psi}\psi + \frac{\lambda}{2} (\dual{\psi}\psi)^2. $$

\section{Specific Non-Standard Spinor Models}
\label{sec:sNSS}

\subsection{Eigenspinors of $\opp{C}$}

In Refs.~\cite{jcap,prd}, Ahluwalia-Khalilova and Grumiller introduced the class of non-standard spinors (is the sense defined above). They constructed these spinors in momentum space from the eigenspinors of the charge conjugation operator and hence called them \emph{Eigenspinoren des LadungsKonjugationsOperators} (ELKOs). The were shown to belong to a non-standard Wigner class~\cite{daRocha:2005ti} and satisfy $(CPT)^2=-\mathbb{I}$. In more mathematical terms, they belong to a wider class of spinorial fields, so-called flagpole spinor fields which corresponds to the class 5 of the Lounesto's classification based on bilinear covariants.

The idea behind ELKOs is attempt to construct a spinor field, $\psi(x)$, from momentum space eigenspinors of the charge conjugation operator, $\lambda(\mathbf{p},e,h)$ say, rather than the $u(\mathbf{p},\sigma)$ and $v(\mathbf{p},\sigma)$ in case of Dirac spinors. The $\lambda(\mathbf{p},e,h)$ are defined by:
\be
\opp{C} \lambda(\mathbf{p},e,h) \equiv -\gamma^{2} \lambda^{\ast}(\mathbf{p},e,h) = e\lambda(\mathbf{p},e,h), \\
\opp{H}(\hat{\mathbf{p}})\lambda(\mathbf{p},e,h) = h\lambda(\mathbf{p},e,h),
\ee
where $e,\, h = \pm 1$ and $\opp{H}(\hat{\mathbf{p}})$ is the dual helicity operator:
\be
\opp{H}(\hat{\mathbf{p}}) = \begin{pmatrix} \sigma\cdot \hat{\mathbf{p}} & 0 \\ 0 & -\sigma\cdot \hat{\mathbf{p}} \end{pmatrix} = \gamma^{0}\gamma^{i}\hat{\mathbf{p}}^i.
\ee
The general free ELKO field is then given by:
\be
\psi(x) = \sum_{\mathbf{p},h} a_{h}(\mathbf{p}) \lambda(\mathbf{p},+1,h)e^{-ip_{\mu}x^{\mu}} + \sum_{\mathbf{p},h} b_{h}^{\dagger}(\mathbf{p}) \lambda(\mathbf{p},-1,h)e^{ip_{\mu}x^{\mu}}. \nonumber
\ee
where $p^{\mu} = (E_{p},\mathbf{p}^{i})$, $E_{p} = \sqrt{\mathbf{p}^2+m^2}$.  It should be noted that the conditions which define the basis spinor $\lambda$, do not do so uniquely (i.e. uniquely up to an overall phase). Instead we have:
\be
\lambda(\mathbf{p}, +1, \pm 1)&=& \bv \pm i e^{i\alpha(p)} \phi_{\pm}(p) \\ \phi_{\mp}(p) \ev, \\
\lambda(\mathbf{p}, -1, \pm 1)&=& \bv \mp i e^{i\alpha(p)} \phi_{\pm}(p) \\ \phi_{\mp}(p) \ev,
\ee
where $\phi_{\pm}(\mathbf{p})$ obey $\sigma\cdot \mathbf{p} \phi_{\pm}(\mathbf{p}) = \pm \phi_{\pm}(\mathbf{p})$, and $\phi^{\dagger}_{a}(\mathbf{p})\phi_{b}(\mathbf{p}) = \delta_{ab}$. It is straightforward to check that any two such spinors will be related by: $\phi_{\pm}(p) = \mp e^{-i\alpha(p)} i\sigma_{2} \phi_{\mp}^{\ast}(p)$ for some $\alpha(p)$, which also features in the definition of $\lambda$, which depends on the phases in the definitions of $\phi_{\pm}$. It can also be checked that $\phi_{\pm}(p) = \mp i e^{\pm i\beta(p)} F(p) \phi_{\mp}(p)$ for some $\beta(p)$. Here $F(p)^2 = 1$ and $F^{\dagger} = F$ and $F(p) = \hat{\mathbf{n}}\cdot \sigma$ where $\hat{\mathbf{n}}(p)$ is a unit vector in the direction of $\mathbf{n}$ which is defined by $\mathbf{n} = \hat{\mathbf{p}} \times \hat{\mathbf{z}}$; here $\mathbf{\hat{z}}$ is a unit vector in some fixed direction. The original definition of ELKOs in Refs.~\cite{jcap,prd} effectively picked $\phi_{\pm}$ so that $\alpha=\beta=0$ and worked in a basis where $\hat{\mathbf{z}} = (0, 0, 1)^{T}$.

To complete the definition of ELKOs one must now find some operator $\opp{P}$, which in momentum space is an odd function of $p^{\mu}$, such that $\opp{P}\psi = \psi$. We note since $a_{h}$ and $b_{h}$ are arbitrary, $\opp{P}$ must commute with the $a_{h}$ and $b_{h}^{\dagger}$. Refs.~\cite{jcap,prd} did not approach the definition of non-standard spinors in the general way that we laid out in the previous section, and so did not explicitly construct $\opp{P}$.  Explicitly constructing $\opp{P}$, however, reveals that the above definition of ELKOs is not Lorentz invariant, see also~\cite{Ahluwalia:2008xi,Ahluwalia:2009rh}. We find that for ELKOs $\opp{P} =\opp{P}_{\rm ELKO}$ where in momentum space:
\be
\opp{P}_{\rm ELKO}(p^{\mu};\hat{\mathbf{z}}) = \begin{pmatrix} 0 & \hat{\mathbf{n}}\cdot \sigma e^{i\alpha -i\beta \sigma\cdot \hat{\mathbf{p}}} \\ \hat{\mathbf{n}}\cdot \sigma e^{-i\alpha + i\beta \sigma\cdot \hat{\mathbf{p}}} & 0 \end{pmatrix}. \nonumber
\ee
It follows that with $\alpha = \beta = 0$ as in Refs.~\cite{jcap,prd}, $\opp{P}_{\rm ELKO}$ simplifies to:
\be
\opp{P}_{\rm ELKO}(p^{\mu};\hat{\mathbf{z}}) = \begin{pmatrix} 0 & \hat{\mathbf{n}}(\mathbf{p},\hat{\mathbf{z}})\cdot \sigma \\ \hat{\mathbf{n}}(\mathbf{p},\hat{\mathbf{z}})\cdot \sigma & 0 \end{pmatrix}.
\ee
It can be checked that modulo the relation ($p^2 =m^2$) there is no other operator $\opp{P}$, with $\opp{P}^2 = \mathbb{I}$ that satisfies the required properties. It is also clear that because $\opp{P}(p)$ depends on both a preferred direction $\hat{\mathbf{z}}$ and $\hat{\mathbf{p}}$ it is not Lorentz invariant. Thus the initial definition of the ELKO basis and hence ELKO field is also not Lorentz invariant as it implicitly assumes the existence of a preferred direction $\hat{\mathbf{z}}$. Additionally when $\hat{\mathbf{p}}=\hat{\mathbf{z}}$, $\mathbf{n} = \mathbf{0}$ and so $\hat{\mathbf{n}}$ and hence $\opp{P}(p)$ is not defined. In the limit $\hat{\mathbf{p}} \rightarrow \hat{\mathbf{z}}$, the limiting value of $\hat{\mathbf{n}}$ depends on the direction of approach. The original definition therefore suffers from a number of issues: it requires a preferred space-like direction, and is ill-defined for momentum pointing along that direction. Similar issues will arise if different (potentially $p$-dependent) values of $\alpha$ and $\beta$ are taken. The Lorentz violation in the definition of ELKOs was not clear in the original papers because, primarily, they did not approach the construct of non-standard spinors in the general covariant manner that was laid out in the previous section.

\subsection{Modified Eigenspinors of $\opp{C}$}

We can make an alternative definition of an ELKO, motivated by more exotic ideas, which is well-defined for all momenta and invariant under rotations, albeit not under boosts. The basis spinors are once again eigen-spinors of the charge conjugation operator. This time however we define the ELKO field by:
\be
\psi(x) = \sum_{\mathbf{p},e} a_{e}(\mathbf{p}) \lambda(\mathbf{p},e,+1)e^{-ip_{\mu}x^{\mu}} + \sum_{\mathbf{p},e} b_{e}^{\dagger}(\mathbf{p}) \lambda(\mathbf{p},e,-1)e^{ip_{\mu}x^{\mu}}.
\ee
This definition has the advantage of being now independent of the phases in the definitions of the $\phi_{\pm}$ two-spinors used to construct the $\lambda(\mathbf{p},e,h)$. It is also straight-forward to check that it is invariant under rotations. This can be seen explicitely by noting that the projection operator, $\opp{P}_{\rm M-ELKO}$, under which $\opp{P}_{\rm M-ELKO} \psi = \psi$ is given in momentum space by simply the dual helicity operator $\opp{H}(\mathbf{p})$:
\be
\opp{P}_{\rm M-ELKO}(p^{\mu}) = \begin{pmatrix} \sigma\cdot \hat{\mathbf{p}} & 0 \\ 0 & -\sigma\cdot \hat{\mathbf{p}} \end{pmatrix} = \gamma^{0}\gamma^{i}\hat{\mathbf{p}}^i. \nonumber
\ee
This operator is manifestly invariant under SO(3) rotations, but also manifestly not invariant under general boosts. However we can write it in a covariant manner by introducing a preferred unit time-like direction $A_{\mu}$ with $A_{\mu}A^{\mu} =1$. We may then choose coordinates so that $A_{\mu} = (1, \mathbf{0})$ and with $p_{s}^{\mu} = (0,\mathbf{p}^{i}) = p^{\mu} - A_{\mu}A{\nu}p^{\nu}$ we have $\hat{p}^{\mu} = (0, \hat{\mathbf{p}}^{i}) = p_{s}^{\mu}(p,A)/ \sqrt{p^{\nu}_s(A,p)p_{s \nu}(A,p)}$. It follows that in this frame: $\opp{P}_{\rm M-ELKO}(p^{\mu}) = \opp{P}_{\rm M-ELKO}(p^{\mu})(p^{\mu};A^{\mu})$ where
\be
\opp{P}_{\rm M-ELKO}(p^{\mu},A^{\mu}) = p_{s \mu} A_{\nu}\gamma^{[\mu}\gamma^{\nu]} = \frac{p_{\mu}A_{\nu}\gamma^{[\mu}\gamma^{\nu]}}{\sqrt{p^2 - (A\cdot p)^2}}.
\ee
We then define $\dual{\lambda} = -\bar{\lambda} \gamma^{\mu}p_{s \mu}$, and by working in the frame where $A_{\mu} = (1,\mathbf{0})^{\rm T}$ find:
\be
\pm\dual{\lambda}(\mathbf{p},e,\pm)\opp{P}_{\pm}^{\rm M-ELKO}\lambda(\mathbf{p},e,\pm) > 0,
\ee
where $\opp{P}_{\pm}^{\rm M-ELKO} = \frac{1}{2}(\mathbb{I} \pm \opp{P}_{\rm M-ELKO})$.

From the above discussion, it is clear that the original ELKO definition breaks Lorentz invariance as, when written in a covariant form, it clearly requires the existence of a preferred space-like direction, $\hat{\mathbf{z}}$. Additionally, the original definition breaks down, even if one takes a limit, for spinors with momentum in the direction of $\hat{\mathbf{z}}$. We have also given a second definition for an ELKO like field. This again breaks Lorentz invariance, but preserved rotational invariance, as it only requires the introduction of a preferred time-like direction $A_{\mu}$. Of course any violation of Lorentz invariance is arguably a serious reason to doubt a theory.

\subsection{Lorentz Invariant Non-Standard Spinors}
\label{sec:LoInNSS}

We have shown above that the original definition of ELKO fields violates Lorentz invariance by introducing a preferred space-like direction. Additionally we noted a modification of this definition could be made, still using dual helicity basis spinors, where there is a frame choice where rotational invariance is preserved, however this required the introduction of a preferred time-like direction, which also violates Lorentz invariance. We may therefore wonder whether there is any reasonable Lorentz invariant definition of projection operator $\opp{P}(p^{\mu})$ which obeys $\opp{P}^2 = \mathbb{I}$ (independently of the field equation $p^2 =m^2$ i.e. without placing any restriction on $p^2$) as well as the other conditions on $\opp{P}$. We know that any such operator must be non-local in position space, since it has been shown that the assumptions of locality and Lorentz invariance imply that the only spin 1/2 field theory is that of Dirac / Majorana spinors. Working in a flat background and in momentum space, Lorentz invariance implies that we cannot introduce any preferred frame-fields. Hence $P(p^{\mu})$ can only be constructed from the Lorentz covariant operators $\mathbb{I}$, $p^{\mu}$ and $\gamma^{\mu}$. Additionally the requirement that $P^2 = \mathbb{I}$ independently of the value of $p^2$ implies that the operator should not depend on the on-shell value of $p^2 (=m^2)$. Additionally we know that $\opp{P}$ must be an odd function of $p^{\mu}$ and $\opp{P} = \oppD{P}$.

Taken together these conditions imply that, up to an arbitrary phase factor, the only choice we can make for $\opp{P}(p^{\mu})$ is:
\be
\opp{P}(p^{\mu})  \equiv \frac{1}{2}(1+i\gamma^{5})\opp{P}_{0}(p^{\mu}) +\frac{1}{2}(1-i\gamma^{5})\dual{\opp{P}}_{0},
\ee
where:
\be
\opp{P}_0 = \opp{p}^{-1}\slashed{p},
\ee
and $\opp{p} = \sqrt{p_{\mu}p^{\mu}}$ with some appropriate choice of branch for the square root's action on negative $p_{\mu}p^{\mu}$.   Thus we have $\opp{P}_0^2 = \mathbb{I}$. We shall see below that with this choice we must take $\dual{\psi} = \bar{\psi}$; it follows that $\dual{\gamma}^{5} = \gamma^{0}\gamma^{5 \dagger} \gamma^{0} = -\gamma^{5}$. Hence the operator $i\gamma^{5}$  is self-dual.  Additionally $\gamma^{5}$ anti-commutes with $\opp{P}_{0}$.  Given this choice of dual, it is straight-forward to check that $\oppD{P}_{0} = \pm \opp{P}_{0}$ depending on the sign of $p_{\mu}p^{\mu}$; hence $\oppD{P}_0$ and $\opp{P}_0$ commute and $\oppD{P}_0^2 = \mathbb{I}$.  Additionally, the dual of $\oppD{P}_0$ is therefore $\opp{P}_0$.

We found that the first condition that $\opp{P}$ must obey is that it is an odd function of momentum.  Since $\opp{P}_0$ is manifestly an odd-function of $p^{\mu}$ and $\opp{P}$ is also.  Secondly we must check that $\opp{P}^2 = \mathbb{I}$. Explicitly:
\be
\opp{P}^2 &=& \frac{1}{4}(1+i\gamma^{5})\opp{P}_0 (1+i\gamma^{5})\opp{P}_0  + \frac{1}{4}(1-i\gamma^{5})\oppD{P}_0 (1-i\gamma^{5})\oppD{P}_0 \\ &&+ \frac{1}{4}(1+i\gamma^{5})\opp{P}_0 (1-i\gamma^{5})\oppD{P}_0 + \frac{1}{4}(1-i\gamma^{5})\oppD{P}_0 (1+i\gamma^{5})\opp{P}_0, \nonumber \\
&=& \frac{1}{4}(1+i\gamma^5)(1-i\gamma^5)\left[ \opp{P}_0^2 + \oppD{P}_0^2 \right] \nonumber \\ &&+ \frac{1}{4} \left[ (1+i\gamma^5)^2 \opp{P}_0 \oppD{P}_0 +(1-i\gamma^5)^2 \oppD{P}_0 \opp{P}_0\right], \nonumber \\
&=& \mathbb{I} + \frac{1}{2} i\gamma^5\left[ \opp{P}_0, \oppD{P}_0\right] = \mathbb{I}. \nonumber
\ee
Here we have used the anti-commutation of $\gamma^5$ with $\opp{P}_0$  $\oppD{P}_0$  and the commutation of $\opp{P}_0$ and $\oppD{P}_0$.  We have also used $\opp{P}_0^2 = \oppD{P}_0^2 = \mathbb{I}$. Thus as require $\opp{P}$ is an odd-function of momentum, and $\opp{P}^2 \equiv \mathbb{I}$ on any spinor.  We also note that since $\dual{\psi} = \bar{\psi}$, $(\dual{\psi}\psi)^{\dagger} = \dual{\psi}\psi$ and the action is real provided the final condition $\opp{P} = \oppD{P}$ (i.e. $\opp{P}$ is self-dual) is satisfied.    We show this explicitly:
\be
\oppD{P} &=& \frac{1}{2} \oppD{P}_0 (1+i\gamma^{5}) + \frac{1}{2} \opp{P}_0 (1-i\gamma^5), \\
&=& \frac{1}{2}(1-i\gamma^{5}) \oppD{P}_0  + \frac{1}{2} (1+i\gamma^5)\opp{P}_0  = \opp{P}, \nonumber
\ee  
where we have used the anti-commutation of $\gamma^{5}$ and $\opp{P}_0$ and the self-dual nature of $i\gamma^{5}$.   

Thus this choice of $\opp{P}$ satisfies all the required properties and is manifestly Lorentz invariant. The appearance of $\opp{p}^{-1}$ factor means that in position space $\opp{P}$ will be non-local, as expected.

In flat position space:
\be
\opp{P} = \frac{1}{2}(1+i\gamma^{5})\opp{P}_{0} +\frac{1}{2}(1-i\gamma^{5})\dual{\opp{P}}_{0}, \qquad \opp{P}_0 = -i \opp{p}^{-1}\slashed{\partial}, \nonumber
\ee
where in position space: $\opp{p}^{-1} = \sqrt{-\partial^{-2}}$; $\partial^{-2}$ is the inverse of $\partial^2$. In a general curved space-time $\opp{p}^{-1}$ is the inverse square root of $\slashed{\nabla}^2$. The definition of what is meant by this square root is fixed by the requirement that in flat-space $\opp{p}^{-1}$ it reduce to $\sqrt{-\partial^2}$.

In curved space-time it is not immediately obvious that the extension of $\opp{p}^{-1}$ commutes with $\slashed{\nabla}$. To prove that this is indeed the case we consider eigenstates of $-i\slashed{\nabla}$ i.e. $-i\slashed{\nabla}\psi_{z}(x^{\mu};s) = -iz \psi_{z}(x^{\mu};s)$ for some $z \in \mathbb{C}$ where $s$ labels the multiplicity. Now $\opp{p}^2 = -\slashed{\nabla}^2$ and so $\opp{p}^2 \psi_{z}(x^{\mu};s) = -z^2 \psi_{z}(x^{\mu};s)$. The operator $\opp{p}^{-1} = \sqrt{-\slashed{\nabla}^2}$ where the square-root requires a choice of branch to make it unambiguous. We right $z = R e^{i\Theta}$ where $R > 0$ and $-\pi < \Theta \leq \pi$. We choose the branch so that $\opp{p} \psi_{z}(x^{\mu};s) = -\varepsilon i z \psi_{z}(x^{\mu};s)$ where $\varepsilon = -1$ for $0 <\Theta \leq \pi$ and $\varepsilon = +1$ for $-\pi < \Theta \leq 0$. Thus in a general space-time the definition of $\opp{p}^{-1}$ is fixed by $\opp{p}^{-1} \psi_{z}(x^{\mu};s) = i\varepsilon z^{-1}\psi_{z}(x^{\mu};s)$. Since $\opp{p}^{-1}$ and $\slashed{\nabla}$ have simultaneous eigenstates (by definition) it is automatic that $\opp{p}^{-1}$ and $\slashed{\nabla}$ commute.Acting on $\psi_{z}(x^{\mu};s)$, $\opp{P}_0\psi_{z}(x^{\mu};s) = \varepsilon \psi_{z}(x^{\mu};s)$ where $\varepsilon = -{\rm sign}({\rm arg} z)$ and $\varepsilon = +1$ if ${\rm arg}z =0$; $-\pi < {\rm arg} z \leq \pi$. 

In general backgrounds:
\be
\opp{P} = \frac{1}{2}(1+i\gamma^{5})\opp{P}_{0} +\frac{1}{2}(1-i\gamma^{5})\dual{\opp{P}}_{0}, \qquad \opp{P}_0 = -i \opp{p}^{-1}\slashed{\nabla} = -i \slashed{\nabla}\opp{p}^{-1}. \label{PDef}
\ee

We have stated above that $\dual{\psi} = \bar{\psi}$. It is now straight-forward to show that this must be the case.  .  We can always write $\dual{\psi} = \bar{\psi} \oppL{D}$ for some operator $\oppL{D}$.  We must first require that $\dual{\psi}\psi$ is real, which in turn requires that $\oppL{D} = \oppD{D}$.  It implies that $\oppL{D} = \oppD = c_0 \mathbb{I} + c_1 i \gamma^{5}$.  We also need $\oppL{P} = \oppDL{P}$. We found that if $\oppL{D} = \mathbb{I}$ then $\oppL{P} = \oppDL{P}$.  For general $\oppL{D}$ we have: $\oppL{P}\oppL{D} = \oppL{D} \oppDL{P}$, hence the reality condition $\opp{P} = \oppD{P}$, or equivalently  $\oppL{P}  = \oppDL{P}$, requires that $\oppL{D}$ commutes with $\oppL{P}$.  Since $\gamma^{5}$ anti-commutes with $\opp{P}_0$, it also anti-commutes with $\opp{P}$ and hence $\oppL{P}$.  It follows that $c_1 = 0$ and we can then normalize $\psi$ so that $c_0 = 1$ and $\oppL{D} = \mathbb{I}$ i.e. $\dual{\psi} = \bar{\psi} = \psi^{\dagger}\gamma^0$.

For a free-field, obeying $\slashed{\nabla}^2 \psi = -m^2 \psi$, $q^{-1} = m^{-1}$ and so $\opp{P} \psi = \psi$ reduces to the Dirac equation. The dynamics of free fields defined in this way are therefore identical to those of Dirac fermions. The differences between the definitions are only apparent for non free fields e.g. those where $V(\dual{\psi}_{+}\psi_{+})$ is a non-linear (e.g. quartic) function of the invariant $\dual{\psi}_{+}\psi_{+}$.

We also note that $\opp{P}$ manifestly commutes $\slashed{\nabla}^2 = -\mathsf{q}^{2}$ and hence
$$(\slashed{\nabla}^2 \psi_{+})_{+} = \slashed{\nabla}^2 \psi_{+} = (\slashed{\nabla}^2 \psi)_{+}.$$

In Appendix \ref{app:quant}, we given a general treatment of the path integral quantization of non-standard spinor models with special reference to the Lorentz invariant model presented above.  The free field, flat space propagator for this theory is:
\be
G_{\rm NSS}(p;m) = \frac{\opp{P}_{+}(p^{\mu})}{p^2-m^2+i\epsilon}.
\ee
One might be worried that the $\opp{p}^{-1}$ term in this propagator (which comes from $\opp{P}_{0}$ in  $\opp{P}(p^{\mu})$) leads to some additional divergences in quantum amplitudes. In Appendix \ref{app:quant:free} and \ref{app:quant:int} we show explicitly that this is not the case.  Simply, this may be understand from the fact that $\opp{p}^{-1}$ does not lead to a new pole in the components of $\opp{p}^{\mu} = (\omega, \mathbf{p})$, since $\opp{p} = \sqrt{\omega-\vert \mathbf{p}\vert}\sqrt{\omega+\vert \mathbf{p}\vert}$ and so the integral of $\opp{p}^{-1}$ with respect to the momenta does not diverge.    It follows that the position space Green's function, $G_{\rm NSS}(x-y)$ (related by Fourier transform to $G_{\rm NSS}(p;m)$) is well defined.   Self-interactions of $\psi$ can be introduced as perturbations about the free field theory. For completeness, we show explicitly, in Appendix \ref{app:quant:int}, that the $\opp{p}^{-1}$ in $G_{\rm NSS}(p;m)$ does not lead to any new divergences in loop integrals and so the quantum theory may be rendered finite by re-normalization in the usual way.

We also find in Appendix \ref{app:quant:free}, that there is a preferred choice of kinetic term.   Specifically one wishes $\opp{P}$ to commute with the free-field Green's function in a general background.  In a general background $\opp{P}$ commutes with $\slashed{\nabla}^2$ but not with $\nabla^2$.  Thus the preferred structure for the action of a Lorentz invariant non-standard spinor, with arbitrary sources $J$ and $\dual{J}$ is:
\be
\mathcal{L}_{\rm NSS} &\equiv& ( \dual{\psi}\overleftarrow{\slashed{\nabla}})(\slashed{\nabla}\psi) - m^2_{\psi}\dual{\psi}\psi - \dual{\chi}\opp{P}_{-}\psi - \dual{\psi}\oppL{P}_{-}\chi + \dual{J}\psi + \dual{\psi}J.
\ee

\section{Energy Momentum Tensor}
\label{sec:emt}

Let us now construct the energy-momentum tensor based on the above actions, not taking into account any of the possible interaction terms. By definition, we need to vary the Lagrangian with respect to the metric $g_{\mu \nu}$. In all previous treatments, when the variation with respect to the metric was computed, the implicit dependence of the connection on the metric was neglected. This happened because in the case of Dirac spinors, one can indeed neglect this contribution as it vanishes identically. Although this relatively well known to experts, we show this calculation explicitly in Appendix~\ref{appdirac}. Therefore, we will now show in detail the derivation of the complete energy-momentum tensor of ELKO spinors. It should be noted however, that one can start with an effective scalar field Lagrangian which contains a mass dependent on the Hubble parameter to reproduce the previous results.

We now derive a formal expression for the energy momentum tensor for the two possible actions $\mathcal{S}_{\psi}^{(1)}$ and $\mathcal{S}_{\psi}^{(2)}$ with Lagrangian density $\mathcal{L}_{\psi-P}^{(1)} = \mathcal{L}_{\psi}^{(1)} + \mathcal{L}_{P}$ and $\mathcal{L}_{\psi-P}^{(2)}=\mathcal{L}_{\psi}^{(2)} + \mathcal{L}_{P}$, where $\mathcal{L}_{\psi}^{(1)}$, $\mathcal{L}_{\psi}^{(2)}$ are given by Eqs.~\eqref{action1} and \eqref{action2} respectively. This derivation is complicated by the presence of the projection action $\mathcal{L}_{P}$, which projects out the `unphysical' modes, it is an odd function of momentum and it has a dependence on $\nabla_{\mu}$ and $\gamma^{\mu}$ therefore on the metric itself complicating the derivation of the energy momentum tensor. Since $P$ is also a non-local operator, its dependence on $g_{\mu \nu}$ is also generally extremely complicated, and this prevents us from finding an general explicit expression for $T^{\mu \nu}_{\psi}$. In some choices of $P$ in certain backgrounds, however, $\delta S_{\psi}/\delta P = 0$ and in these cases we can give an explicit expression for $T^{\mu \nu}_{\psi}$.

\subsection{Variation of $\nabla_{\mu}$}
It is important to remember when calculating $T_{\mu \nu}$ that $\nabla_{\mu}$ also depends on the gamma matrices $\gamma_{\mu }$ through $\Gamma_{\mu \nu}$. Thus:
\be
\frac{\delta \nabla_{\rho}\psi}{\delta g_{\mu \nu}} = -\frac{\delta \Gamma_{\rho}}{\delta g_{\mu \nu}}\psi, \\
\frac{\dual{\psi}\delta \overleftarrow{\nabla}_{\rho}}{\delta g_{\mu \nu}} = \dual{\psi} \frac{\delta \Gamma_{\rho}}{\delta g_{\mu \nu}}.
\ee
Now $\Gamma_{\mu} = i \omega_{\mu}{}^{ab} f_{ab}/4$ and the spin connection depends on $e_{\mu}{}^{a}$ and hence $g_{\mu \nu}$. We calculate $\delta \omega_{\mu}{}^{ab}$ in a local inertial frame (LIF) where $e_\mu^{a} = \delta_{\mu}^a$ and so $\Gamma_{\mu \nu}^{\rho} = 0$.
In a LIF, we find:
\be
f_{ab}\delta \omega_{\mu}{}^{ab} = -f_{ab} e^{\nu b}e_{\rho}^{a}\partial_{\mu}\left[ e_{c}^{\rho}\delta e_{\nu}^{c}\right] + f_{ab} e^{a}{}_{\nu}e^{b \sigma}\delta\Gamma_{\mu \sigma}^{\nu}.
\ee
Now in a LIF:
\be
f_{ab}e^{a}{}_{\nu}e^{b \sigma}\delta \Gamma_{\mu \sigma}^{\nu} = \frac{1}{2}f_{ab}e^{\nu a}e^{\sigma b}\left[\delta g_{\mu \nu,\sigma} + \delta g_{\nu \sigma,\mu} - \delta g_{\mu \sigma,\nu}\right] \nonumber = f^{\nu \rho}\delta g_{\mu \nu,\rho}. \nonumber
\ee
To move to a general frame we promote partial derivatives to covariant derivatives and have:
\be
f_{ab}\delta \omega_{\mu}{}^{ab} = f^{\nu \rho} \nabla_{\mu}\left[ e_{\rho c}\delta e_{\nu}^{c}\right] + f^{\nu \rho} \nabla_{\rho}\left[\delta g_{\mu \nu}\right]. \label{domega}
\ee
where we have used $f^{ab} = -f^{ba}$ and defined $f^{\mu \nu} = e^{\mu a}e^{\nu b} f_{ab}$. Now $\delta g_{\mu \nu} = 2e_{(\mu}^{a}\delta e_{\nu) a}$ and the first term in Eq.~\eqref{domega} depends only on $e_{[\mu}\delta e_{\nu]}^{a}$ and which is independent of the variation in $g_{\mu \nu}$. Hence just varying $g_{\mu \nu}$:
\be
\delta \Gamma_{\rho} = \frac{i}{4}\delta_{\rho}^{(\mu}f^{\nu) \sigma}\nabla_{\sigma} \delta g_{\mu \nu}.
\ee

\subsection{Variation of $\gamma^{\mu}$}
Now $\gamma^{\mu} = e^{\mu}_{a}\gamma^{a}$ and so:
\be
\delta \gamma^{\mu} = -e^{\mu}_{b}e^{\nu}_{a}\delta e_{\nu}^{b} \gamma^{a},
\ee
and so just varying $g_{\mu \nu}$:
\be
\delta \gamma^{\rho} = -\frac{1}{2} \gamma^{(\nu} g^{\mu )\rho} \delta g_{\mu \nu}.
\ee

\subsection{Variation of $\mathcal{L}_{\psi}^{i}$}
The total Lagrangian is $\mathcal{L}_{\psi-P}^{(i)} = \mathcal{L}_{\psi}^{(i)} + \mathcal{L}_{P}$. We consider the variation of the projection term $\mathcal{L}_{P}$ separately. Here we simply calculate $T^{(i) \mu \nu}_{\psi} = \left[-2/\sqrt{-g}\right]\delta \mathcal{S}_{\psi}^{(i)}/\delta g_{\mu \nu}$ where the action $\mathcal{S}_{\psi}^{(i)}$ is the integral of $\mathcal{L}_{\psi}^{(i)}(\psi,\dual{\psi})$. This action is independent of $P$.

We find:
\be
T^{(1) \mu \nu}_{\psi} = \dual{\psi} \overleftarrow{\nabla}^{(\mu} \gamma^{\nu)} \slashed{\nabla}\psi + \dual{\psi} \overleftarrow{\slashed{\nabla}} \gamma^{(\mu}\nabla^{\nu)}\psi - g^{\mu \nu} \mathcal{L}_{\psi}^{(1)} + \nabla_{\rho} J^{\rho \mu \nu}_{(1)},
\ee
where the last term comes from the variation of $\Gamma_{\mu}$ and is equal to:
\be
J^{\mu \nu \rho}_{(1)} = -\frac{i}{2}\left[\dual{\psi} \overleftarrow{\slashed{\nabla}} \gamma^{(\mu}f^{\nu)\rho}\psi + \dual{\psi} f^{\rho (\mu}\gamma^{\nu)} \slashed{\nabla}\psi\right].
\ee

For the second action we have:
\be
T^{(2) \mu \nu}_{\psi} = 2\dual{\psi} \overleftarrow{\nabla}^{(\mu}\nabla^{\nu)}\psi - g^{\mu \nu} \mathcal{L}_{\psi}^{(2)} + \nabla_{\rho} J^{\rho \mu \nu}_{(2)},
\ee
where again the last term comes from the variation of $\Gamma_{\mu}$ is given by:
\be
J^{\mu \nu \rho}_{(2)} = -\frac{i}{2}\left[\dual{\psi} \overleftarrow{\nabla}^{(\mu} f^{\nu)\rho}\psi + \dual{\psi} f^{\rho (\mu} \nabla^{\nu)}\psi\right].
\ee

\subsection{Variation of $\mathcal{L}_{P}$}

We now focus on the variation of $\mathcal{L}_{P}$ with respect to $\opp{P}$. We have:
\be
\delta \mathcal{L}_{P} = \frac{1}{2}\dual{\chi} (\delta \opp{P}) \psi + \frac{1}{2} \dual{\psi} (\delta \opp{P}) \chi.
\ee
Now $\opp{P}^2 = \mathbb{I}$ implies $\opp{P}_{\pm}\delta \opp{P} = \delta \opp{P} \opp{P}_{\mp}$ and using $\psi = \opp{P}_{+}\psi$, and dropping an irrelevant surface term it is straightforward to check that the above variation reduces to:
\be
\delta \mathcal{L}_{P} = \frac{1}{2}\dual{\chi}_{-} (\delta \opp{P}) \psi_{+} + \frac{1}{2} \dual{\psi}_{+} (\delta \opp{P}) \chi_{-}.
\ee
We see that generally the variation of $\mathcal{L}_{P}$ with respect to $\opp{P}$ and hence $g_{\mu \nu}$ only vanishes if $\dual{\chi}_{-} = \chi_{-} = 0$. To see what this requires, we define the differential operators $\opp{L}_{\psi}$ and $\oppDL{L}_{\psi}$ by:
\be
\frac{\delta \mathcal{S}_{\psi}}{\delta \dual{\psi}} = -\opp{L}_{\psi} \psi, \qquad
\frac{\delta \mathcal{S}_{\psi}}{\delta \psi} = -\dual{\psi} \oppDL{L}_{\psi}.
\ee
The field equations are then $(\opp{L}_{\psi}\psi)_{+} = (\dual{\psi}\oppDL{L}_{\psi})_{+} = 0$ and
\be
\chi_{-} = -(\opp{L}_{\psi}\psi)_{-}, \qquad \dual{\chi}_{-} = -(\dual{\psi}\oppDL{L}_{\psi})_{-}.
\ee
Since, $\psi_{-} = \dual{\psi}_{-} = 0$ we have:
\be
\chi_{-} = \frac{1}{2}\opp{R}_{P}\psi, \qquad \dual{\chi}_{-} = \frac{1}{2}\dual{\psi} \oppDL{R}_{P},
\ee
where
$$
\opp{R}_{P} = \left[\opp{L}_{\psi},\opp{P}\right] = 2[\opp{P}_{-},\opp{L}{\psi}],
$$
and $ \oppDL{R}_{P}$ is the dual operator. It follows that in general $\chi_{-} = \dual{\chi}_{-} = 0$ requires $\opp{R}_{P} = 0$, i.e. $\opp{P}$ must commute with the field equation operator $\opp{L}_{\psi}$. For general $\opp{P}$ we do not expect (in a non-flat background) $\opp{P}$ to commute with either $\slashed{\nabla}^2$ or $\nabla^2$, and so expect $\opp{R}_{P} \neq 0$. We noted that that if we take Lorentz invariant definition of $\opp{P}$ given in \S \ref{sec:LoInNSS} , then we do have $[\opp{P}, \slashed{\nabla}^2]=0$. Hence if we take the action to be $\mathcal{L}_{\psi-P}^{(1)}$, then $\opp{L}_{\psi} = \slashed{\nabla}^2 + V^{\prime}(\dual{\psi}_{+}\psi_{+})$ and:
\be
\opp{R}_{P} = \left[V^{\prime}(\dual{\psi}_{+}\psi_{+}), P\right] = -V^{\prime \prime}(\dual{\psi}_{+}\psi_{+}) \mathcal{P}(\dual{\psi}_{+}\psi_{+}),
\ee
where we define $\mathcal{P}(\alpha)$ acting on c-numbers $\alpha$ by: $\left[\opp{P},\alpha\right] \equiv \mathcal{P}(\alpha)$. In general then, even with a Lorentz invariant choice of $\opp{P}$, we do not have $\opp{R}_{P} = 0$ unless $V^{\prime \prime} = 0$, in which case the theory reduces to that of a Dirac spinor, or if our solution has such symmetries as to ensure $\mathcal{P}(\dual{\psi}_{+}\psi) = 0$.

We do not attempt to calculate $\delta \opp{P} / \delta g_{\mu \nu}$, instead we merely note that, in a general background, this dependence of $\opp{P}$ on $g_{\mu \nu}$ results in an additional contribution to the energy-momentum tensor which we write as $T^{\mu \nu}_{P}$. In some backgrounds, it may be that the symmetries of the solution imply that $\delta \opp{P}$ vanishes for small changes, $\delta g_{\mu \nu}$, in the metric. Such cases represent another way in which $T^{\mu \nu}_{P}$ could vanish.

\section{ELKO Cosmology}
\label{sec:ELKOCos}

In the previous section we attempted to calculate the energy momentum tensor for non-standard spinors. The presence of the operator $P$, which in general has a complicated dependence on the metric, prevented us from explicitly evaluate $T_{\mu \nu}^{\psi}$ except in circumstances where $[L_{\psi},P]=0$ in which case the variation of the $P$ dependent term vanishes on-shell.

In general backgrounds, and for general $P$ it is therefore difficult to make much progress. We therefore begin by focusing on the relatively simple background of a flat FRW spacetime with line element:
\be
\dd s^2 = \dd t^2 - a^2(t) \dd \mathbf{x}^2.
\ee
In this background:
\be
\Gamma_{t} = 0,\qquad \Gamma_{x^{i}} = -\frac{\dot{a}}{2}\gamma^{0}\gamma^{i}
\ee

\subsection{Comments on previous work}

We noted that the definition of the ELKO field theory is not Lorentz invariant since it requires a preferred directions. This casts doubt on the validity of the model and its usefulness for cosmology. Let us emphasize that in general an explicit and complete expression of the energy-momentum tensor is difficult to find since it is expected to include a term from the variation of the $P$ operator.

Previous studies of ELKO cosmology have, however, side-stepped this issue by simply studying the cosmology of a theory defined by that action:
\be
\mathcal{L}_{\rm cosmo} = \frac{1}{2} \dual{\psi} \overleftarrow{\nabla}_{\mu} \nabla^{\mu}\psi - V(\dual{\psi}\psi),
\ee
which is equivalent $\mathcal{L}_{\psi}^{(2)}$ up to a different normalization of the kinetic term (there is an additional factor $1/2$). This is just a different normalization of the spinor field and does not alter the physics. The additional projection term, $\mathcal{L}_{P}$, did not feature in previous studies of ELKO cosmology. With ELKOs, all Lorentz violating terms are located in $\mathcal{L}_{P}$ and it is this term that creates the problems in deriving an explicit expression for the energy-momentum tensor. Analyzing a NSS without any projection term is therefore more straight-forward than a NSS with $\mathcal{L}_{\rm P}$.  This said, strictly speaking, the $\mathcal{L}_{\rm P}$ is \emph{not} optional since it is required to project out ghost modes which would otherwise result in an unstable quantum theory. Nonetheless, one may treat the action $\mathcal{L}_{\rm ELKO-cosmo}$ classically and study its cosmology. Since $P$ does not appear in this action, previous studies have not truly addressed ELKO cosmology but simply the cosmology of an unconstrained spinor with Klein-Gordon action.

We found in the previous section that the energy momentum tensor of an unconstrained NSS cosmology. Adjusting this for the the different kinetic term normalization we have:
\be
T^{\mu \nu}_{\rm cosmo} &=& \dual{\psi} \overleftarrow{\nabla}^{(\mu}\nabla^{\nu)}\psi - g^{\mu \nu} \mathcal{L}_{\rm cosmo} + \frac{1}{2}\nabla_{\rho} J^{\mu \nu \rho}, \nonumber \\
J^{\mu \nu \rho} &=& -\frac{i}{2}\left[\dual{\psi} \overleftarrow{\nabla}^{(\mu}f^{\nu)\rho}\psi + \dual{\psi} f^{\rho (\mu} \nabla^{\nu)}\psi\right]. \nonumber
\ee
The $\nabla_{\rho} J^{\mu \nu \rho}$ term in $T^{\mu \nu}_{\rm cosmo}$ did not appear in previous studies of ``ELKO'' cosmology, since they did not take into account the variation of the spin-action with respect to the metric. We therefore briefly re-derive the cosmology of these models with the corrected energy momentum tensor.

We make the definition $\psi = \varphi \xi$ where $\xi$ is a constant spinor. Now in principle we can have $\dual{\xi}\xi > 0$, $\dual{\xi}\xi = 0$ or $\dual{\xi}\xi < 0$ with $\xi \neq 0$. However the last do possibilities will result in non-positive energy solutions i.e. ghosts, and we should properly therefore concentrate on the non-ghost solutions with $\dual{\xi}\xi > 0$, and by fixing the definition of $\varphi$, we have $\dual{\xi}\xi = 1$.

We note that $J^{\mu \nu \rho} = \tilde{J}^{(\mu \nu) \rho}$ where:
\be
\tilde{J}_{\mu}{}^{\nu \rho}e_{\nu}^{a}e_{\rho}^{b} \equiv J_{\mu}^{ab} &=& -\frac{i}{2}\left[\dual{\psi} \overleftarrow{\nabla}_{\mu} f^{ab} \psi - \dual{\psi} f^{ab} \nabla_{\mu} \psi\right], \nonumber \\
&=& -\frac{i \varphi^2}{2}\left[ \dual{\xi} \Gamma_{\mu} f^{ab} \xi + \dual{\xi} f^{ab} \Gamma_{\mu} \xi\right]. \nonumber
\ee
It is straight-forward to see that $\tilde{J}_{\mu}^{ab} = -\tilde{J}_{\mu}^{ba}$ and since $\Gamma_{t} = 0$ that $\tilde{J}_{t}^{ab} = 0$.  

Now $f^{0j} = i \gamma^{0}\gamma^{j} = i {\rm diag}(\sigma^{j}, -\sigma^{j})$.  Thus:
\be
\tilde{J}_{x^{i}}^{0j} = -\tilde{J}_{x^{i}}^{j0} =  -\frac{\varphi^2 \dot{a}}{4} \left[ \dual{\xi}\left( \gamma^{0}\gamma^{i} \gamma^{0}\gamma^{j} + \gamma^{0}\gamma^{j} \gamma^{0}\gamma^{i}\right)\xi\right] = -\frac{\varphi^2 \dot{a}}{2} \delta_{i}{}^{j}. \nonumber
\ee
Now $f^{jk} = \epsilon_{ljk} {\rm diag}(\sigma^{l},\sigma^{l})$.  Thus $\left(\gamma^{0}\gamma^{i} f^{jk} +f^{jk} \gamma^{0}\gamma^{i}\right) = 2 \epsilon_{ijk}\gamma^{5}$ where $\gamma^{5} = {\rm diag}(\mathbb{I}_{2\times 2}, -\mathbb{I}_{2\times 2})$.  It follows that:
\be
\tilde{J}_{x^{i}}^{jk} &=& \frac{\varphi^2 \dot{a} i}{4} \left[ \dual{\xi} \left(\gamma^{0}\gamma^{i} f^{jk} +f^{jk} \gamma^{0}\gamma^{i}\right) \xi \right] = \frac{\varphi^2 \dot{a}}{2} i\epsilon_{ijk}\dual{\xi} \gamma^{5} \xi. \nonumber
\ee
It follows that the only non-vanishing component of $J^{\mu \nu \rho}$ is 
\be
J^{x^{i} t x^{j}} = J^{0 x^{i} x^{j}} = \frac{\varphi^2 \dot{a}}{4a^3} \delta^{i}{}_{j}, \qquad J^{x^{i} x^{j} t} = -\frac{\varphi^2 \dot{a}}{2a^3} \delta^{i}{}_{j}. \nonumber
\ee

Recall that the contribution of the current to the energy-momentum tensor is $\frac{1}{2} \nabla_{\rho} J^{(\mu \nu) \rho}$, thus we define the symmetric tensor $F^{\mu \nu} = F^{(\mu \nu)} = \frac{1}{2}\nabla_{\rho} J^{\mu \nu \rho}$ and find for its non-vanishing components
\be
 F^{t}_{t} &=& \frac{3\dot{a}^2}{4a^2} \varphi^2, \\
 F^{x_{i}}_{x_{j}} &=& \frac{1}{4 a^{2}} \delta_{ij} \frac{{\rm d}}{{\rm d}t} \left[ a^{2}{\dot{a}}{a} \varphi^2\right].
\ee
Therefore, the complete energy-momentum tensor is
\be
T^{t}_{t} &=& \frac{1}{2}\dot{\varphi}^2 + V(\varphi^2) + \frac{3\dot{a}^2}{8a^2} \varphi^2, \\
T^{x_{i}}_{x_{j}} &=& \delta^{i}_{j} \left\lbrace \frac{3\dot{a}^2}{8a^2} \varphi^2+ V(\varphi^2) - \frac{1}{2}\dot{\varphi}^2 \right. \left.+ \frac{1}{4}\left[\frac{\dot{a}}{a} \varphi^2\right]_{,t}\right\rbrace.
\ee

We define energy density, $\rho_{\psi}$, and pressure, $p_{\psi}$, via the diagonal components of the energy-momentum tensor $T^{\mu}{}_{\nu} = {\rm diag}(\rho_{\psi}, -p_{\psi}, -p_{\psi},-p_{\psi})$. Hence:
\be
 \rho_{\varphi} &=& \left[\frac{1}{2}\dot{\varphi}^2 + V(\varphi^2)\right] + \frac{3}{8} H^2 \varphi^2, \\
 p_{\varphi} &=& \left[\frac{1}{2}\dot{\varphi}^2 - V(\varphi^2)\right] - \frac{3}{8} H^2 \varphi^2 - \frac{1}{4}\dot{H} \varphi^2 - \frac{1}{2}H \varphi \dot{\varphi}. \nonumber
 \ee

One can now easily check that $\dot{\rho}_{\varphi} + 3H(\rho_{\varphi} + p_{\varphi}) = 0$ implies, as it should, that the field equation for $\varphi$ is
\be
 \ddot{\varphi} + 3H \dot{\varphi} + 2V^{\prime}(\varphi^2)- \frac{3}{4}H^2 \varphi = 0.
 \label{elkoeom}
\ee

Let us consider now the acceleration equation which contains the usual term $\rho_{\psi}+3p_{\psi}$, we find
\be
\rho_{\varphi} + 3p_{\varphi} = \left[2\dot{\varphi}^2 - V(\varphi^2)\right] - \frac{3}{4}\left(H^2\varphi^2 + \dot{H}\varphi^2 + 2H \varphi \dot{\varphi}\right).
\ee
The Friedman equation with matter source, $\rho_{\rm matter}$ now reads
\be
H^2 = \frac{8\pi G}{3(1- \pi G \varphi^2)} \left[ \frac{1}{2}\dot{\varphi}^2 + V(\varphi^2) + \rho_{\rm matter}\right].
\ee
This form of writing the Friedman equation has a particularly nice interpretation. Namely, the presence of am ``ELKO'' (i.e. an NSS without projection operator term) modifies the effective gravitational coupling constant with $G \rightarrow G_{\rm eff} = G/(1-\pi G \varphi^2)$. This in turn places a simple limit on the maximum value of $\varphi$, $\varphi < 1/\sqrt{\pi G} = 2\sqrt{2} M_{\rm Pl}$, where $M_{\rm Pl} = 1/\sqrt{8\pi G}$ is the reduced Planck mass. However since we have not included a projection term in the Lagrangian, it is not clear to what extent, if at all, such a cosmology can be realized with a Lorentz invariant and ghost-free NSS spinor model.

\subsection{Dimensionless representation of field equations}

Consider a setting in which we only have ELKO spinor fields coupled minimally to gravity, this means we neglect all possible interaction terms. We will now formulate the field equations as an autonomous system of two differential equations. We define $u = \pi G \dot{\varphi}$ and $v = \pi G\varphi^2$ and $V(\varphi^2) = f(v)/(\pi G)^2$. Moreover, we introduce a new time coordinate $\tau = t/\sqrt{\pi G}$ and $h(u,v) = H(t)\sqrt{\pi G}$. Then, for the field equations we find
\begin{align}
 u_{\tau} &= -3h(u,v)u + \frac{3h^2(u,v)}{4}\sqrt{v} - 2f_{,v}(v)\sqrt{v},
 \label{dyn1}\\
 v_{\tau} &= 2\sqrt{v} u,
 \label{dyn2}
\end{align}
where the function $h(u,v)$ is given by
\begin{align}
 h(u,v) = 2\sqrt{\frac{u^2 + 2f(v)}{3(1-v)}}.
 \label{dyn3}
\end{align}

Let us analyze these equations from a dynamical systems point of view for the moment. The critical points of the system are obtained by solving $u_\tau = 0$ and $v_\tau = 0$ for $u$ and $v$. The equation $v_\tau = 0$ is satisfied if either $u=0$ or $v=0$. Thus, we now need to solve the other equation $u_\tau = 0$ for these two cases. Thus, we find three conditions when critical points can exist
\be
 &A:\qquad &u=0,\quad v=0,\\
 &B:\qquad &u=0,\quad f'(v) = f(v)/(1-v),\\
 &C:\qquad &u^2=-f(0),\quad v=0.
\ee
The critical point $A$ always exists and corresponds to $\dot{\varphi}=\varphi=0$. The existence of the two other points depends on the function $f$ and thus on the chosen potential of the ELKO field. Point $C$ exists provided $f(0) \leq 0$, this means that a canonical mass of a quartic self interaction term would yield a critical point identical to point $A$. The most interesting is point $B$ since it depends on the form of the entire function on the positive half line. Note that the equation $f'(v) = f(v)/(1-v)$ can in principle have infinitely many solutions. For example, $f(v)=c/(1-v)$ solve this equation for all values of $v$ and in that case we would encounter a critical line. The function $f(v) = \alpha v + \beta/2\, v^2$ on the other yields up to solution depending on the values of $\alpha$ and $\beta$.

\subsection{De Sitter Solutions}
\label{subELKOdeSitter}

Based on this discussion we can have a closer look at de Sitter type solutions. A de Sitter phase is characterized by $h = h_{0} \neq 0$ where $h_0$ is a constant. Let us for the moment denote $c^2 = 3h_{0}^2/4$. Now Eq.~\eqref{dyn3} implies
\be
 u^2 &=& c^2(1-v) - 2f(v), \\
 \Rightarrow 2u u_{\tau} &=& -(c^2 + 2f_{,v}) v_{\tau} = -(c^2 + 2f_{,v})2\sqrt{v} u, \nonumber
\ee
where we used Eq.~\eqref{dyn2}. This latter equation can be satisfied by either $u \equiv 0$ or we have $u_{\tau} = -(a^2+2f_{,v})\sqrt{v}$.

\paragraph{`Fast Roll' de Sitter Solutions $u \neq 0$:}
With $u \neq 0$, the right hand side of Eq.~\eqref{dyn1} gives
\be
 -(c^2+2f_{,v})\sqrt{v} = u_{\tau} = -2 \sqrt{3} c u + (c^2 - 2f_{,v}(v))\sqrt{v},
\ee
from which we find
\be
 2c^2 \sqrt{v} = 2\sqrt{3}c u = 2\sqrt{3}c \frac{v_{\tau}}{2\sqrt{v}},
\ee
where we used Eq.~\eqref{dyn2} in the last step. Thus, $v$ is a solution of the simple differential equation $v_\tau = h_0 v$ and so $v=v_0 e^{h_0 \tau}$ and therefore $f(v) = 3h_{0}^2/8 - h_{0}^2 v/2$. Note that this corresponds to the solution discussed in Subsection~\ref{subELKOdeSitter}.

\paragraph{`Slow / no roll' de Sitter solutions:}
Another possible de Sitter solutions exist for $u \equiv 0$. In that case we have $u_{\tau} = 0$ and so for $v \neq 0$ must require $v= \bar{v} = \bar{v}(\tau)$ with
\be
 f_{,\bar{v}}(\bar{v}) = \frac{3h^2_0}{8}, \qquad
 f(\bar{v}) = \frac{3h_{0}^2 (1-\bar{v})}{8}.
\ee
Thus we need $\bar{v} = v_{0} ={\rm const}$ where
\be
f_{,\bar{v}}(v_0) = \frac{f(v_0)}{1-v_{0}}.
\ee

These solutions are stable with respect to small homogeneous perturbations if
\be
 f_{,\bar{v}\bar{v}}(v_0) > \frac{2f_{,v}(v_0)}{1-v_0} = \frac{2f(v_0)}{(1-v_0)^2} = \frac{2f^2_{,v}(v_0)}{f(v_0)}.
\ee
Thus, stability requires
\be
 f(v_0) > 0, \qquad
 \frac{(1-v_0)f_{,v}(v_0)}{f(v_0)} = 1, \\
 \frac{f_{,vv}(v_0)f(v_0)}{2f^2_{,v}(v_0)} > 1. \nonumber
\ee
Whilst the first two conditions are straightforward to satisfy, the last is more difficult and imposes restrictions on the form of $f$. As above, let us restrict ourselves to $f(v) = \alpha v + \beta v^2/2$ then the last condition is never satisfied when the first two conditions hold. Recall that this choice of $f$ corresponds to a potential with canonical mass term and quartic self interaction

Solutions with $f(v) = f_0 + \alpha v + \beta v^2/2$ do, however, exist provided certain conditions on $f_0/\alpha$ and $\alpha/beta$ hold. In this case, as opposed to the standard scalar field scenario, $\Phi$ is held up the potential by $H^2$ and so one is not actually at a minimum of $V(\dual{\lambda}\lambda)$.

If there is a minimum of the potential at $\dual{\lambda}\lambda = 0$, then one does, however, get de Sitter type solutions. This could be interesting if $f(0) < f(v_0)$, $f_{,v} (0) > f(0)$ (and the other conditions given above hold when $v=v_0$) as then the de Sitter solution at $v=v_0$ is only actually meta-stable and one would expect it to decay via tunneling to the solution at $v=0$.

\subsection{Conformal couplings}

Let us now briefly outline the case when a conformal coupling $-\beta/2\, \dual{\lambda}\lambda \mathcal{R}$ is taken into account. Then the energy momentum tensor becomes
\be
 T^{\mu \nu} &=& \nabla^{(\mu}\dual{\psi} \nabla^{\nu)} \psi - \frac{1}{2} g^{\mu \nu} \nabla^{\rho} \dual{\psi} \nabla_{\rho} \psi \\ &&+ \nabla_{\rho} J_{\rm elko}^{(\mu \nu)\rho} + g_{\mu\nu}V(\dual{\psi}\psi)+ \beta\, \nabla^{\mu}\nabla^{\nu} (\dual{\psi}\psi) - \beta\, \square (\dual{\psi}\psi) g^{\mu \nu}. \nonumber
\ee
The Einstein field equations are modified to
\be
 (1-8\pi G \beta \dual{\lambda}\lambda) G^{\mu \nu} = -8\pi G \left[T^{\mu \nu}_{\rm elko} + T^{\mu \nu}_{\rm matter}\right],
\ee
where the matter energy momentum tensor is defined as usual
\be
 T^{\mu \nu}_{\rm matter} = -2\frac{\delta (\sqrt{-g} \mathcal{L}_{\rm matter})}{\delta g_{\mu \nu}}.
\ee

As before, we consider the case where the cosmological ELKO spinor is given in the form $\psi = \varphi(t) \xi$ where $\xi$ is a constant spinor satisfying $\dual{\xi} \xi = 1$. In this case $\varphi$ obeys
\be
 \ddot{\varphi} + 3H\dot{\varphi} + \left(2V_{,\varphi^2} + \beta R - \frac{3}{4}H^2\right) \varphi = 0,
\ee
which reduces to Eq.~\eqref{elkoeom} in the limit $\beta \rightarrow 0$. Moreover, the effective energy density of the conformally coupled ELKO field is given by
\be
 \rho_{\rm elko} = \frac{1}{2}\dot{\varphi}^2 + V(\varphi^2) + \frac{3}{8}H^2 \varphi^2 - 6\beta H \varphi \dot{\varphi}.
\ee
The modified Friedman equation now take the form
\be
3(1-\beta 8\pi G \varphi)H^2 = 8\pi G \left[\rho_{\rm elko} + \rho_{\rm matter}\right].
\ee
It is clear that de Sitter type solutions will exist also in the conformally coupled case since there is now an additional degree of freedom and a more interesting coupling between the matter and the geometry. Since this scenario has not been studied yet for the ELKO spinor field, we expect a variety of interesting results to emerge.

\section{Cosmology of Lorentz Invariant NSS}
\label{sec:NSSCos}

We now focus on the cosmology of non-standard spinors with the Lorentz invariant form of $P$:
\be
\opp{P} = \frac12(1+i\gamma^5) \opp{P}_0 + \frac12(1-i\gamma^5) \oppD{P}_0, \qquad \opp{P}_0 = -i \opp{p}^{-1}\slashed{\nabla}.
\ee
We also focus on the action given by $L_{\psi}^{(1)}$ (i.e with the kinetic term $\dual{\psi}\overleftarrow{\slashed{\nabla}}\slashed{\nabla}\psi$) since the resulting field equation operator $\slashed{\nabla}^2$ commutes with $P$. With this choice of kinetic term we do not find any modification to the effective gravitational constant. However, in this case when $V^{\prime} = m^2$ we recover standard Dirac spinors both in curved and flat space and so we can be confident that the free-field theory is ghost-free.

In flat FRW backgrounds $\Gamma_{0} = 0$ and $\Gamma_{x^{i}} = 3\dot{a}/2 \gamma^{i}\gamma^{0}$ so $-\gamma^{\mu}\Gamma_{\mu} = 3\dot{a}/2a$. It follows that:
\be
\slashed{\nabla} \psi = a^{-3/2}\slashed{\partial} (a^{3/2}\psi). \label{FRWnabla}
\ee
Hence:
\be
\opp{P} \psi = a^{-3/2} \opp{P}^{\rm flat} \left( a^{3/2} \psi\right),
\ee
where $\opp{P}^{\rm flat} = (1+i\gamma^5) \opp{P}_0^{\rm flat} + (1-i\gamma^5) \oppD{P}_0^{\rm flat}$ where $\opp{P}^{\rm flat}_0 = -i \opp{p}^{-1}\slashed{\partial}$ and $\opp{p}^{-1}$ is $1/\sqrt{-\partial^2}$. 

The total Lorentz invariant NSS spinor action is:
\be
S_{\psi} = \int \sqrt{-g}\dd^4 x \left[\bar{\psi} \overleftarrow{\slashed{\nabla}} \slashed{\nabla}\psi - V(\bar{\psi}\psi)\right. \left.- \bar{\chi}\opp{P}_{-}\psi - \bar{\psi}\oppL{P}_{-}\chi\right]. \nonumber
\ee
Using the relation between $\slashed{\nabla}$ and $\slashed{\partial}$ and between $\opp{P}$ and $\opp{P}_{\rm flat}$ in an FRW background gives:
\be
S_{\psi} = \int \dd^3 x\int \dd t \left[\bar{\tilde{\psi}} \overleftarrow{\slashed{\partial}} \slashed{\partial}\tilde{\psi} - a^3 V(a^{-3}\bar{\tilde{\psi}}\tilde{\psi})\right. \left.+ \bar{\tilde{\chi}}\opp{P}_{-}^{\rm flat}\tilde{\psi} + \bar{\tilde{\psi}}\oppL{P}^{\rm flat}_{-}\tilde{\chi}\right]. \nonumber
\ee
where $\tilde{\psi} = a^{3/2}\psi$. Varying the full action with respect to $a$ and defining $\tilde{\Phi} = \bar{\tilde{\psi}}\tilde{\psi}$, and $\Phi = a^{-3}\tilde{\Phi}$, gives:
\be
3H^2 + 2\dot{H} = \kappa \left[V(\Phi)- V^{\prime}(\Phi) \Phi\right].
\ee
Assuming that $\psi = \psi(t)$, we find that $\oppD{P} = \opp{P}$ and so $\opp{P}=\opp{P}_0$ and $\opp{P}^{\rm flat} = \opp{P}^{\rm flat}_0$.

The field equations for $\tilde{\psi}$ and $\bar{\tilde{\psi}}$ are then:
\be
\tilde{\psi} = \tilde{\psi}_{+}, \qquad \bar{\tilde{\psi}} = \bar{\tilde{\psi}}_{+}, \nonumber \\
\ddot{\tilde{\psi}} + V^{\prime}(\Phi)\tilde{\psi} = \tilde{\chi}_{-}, \nonumber \\
\ddot{\bar{\tilde{\psi}}} + V^{\prime}(\Phi) \bar{\tilde{\psi}}_{+} = \bar{\tilde{\chi}}_{-}. \nonumber
\ee
We define $\tilde{\Psi} = \dot{\bar{\tilde{\psi}}}\dot{\tilde{\psi}}$ and $\Psi = a^{-3}\tilde{\Psi}$. We then have the coupled equations:
\be
\ddot{\tilde{\Phi}} &=& 2\left[\tilde{\Psi} - V^{\prime}(\Phi)\tilde{\Phi}\right], \label{eq:PhiDDot} \\
\dot{\tilde{\Psi}} &=& -V^{\prime}(\Phi) \dot{\tilde{\Phi}}. \label{eq:PsiDot}
\ee

In an FRW background $\psi = \psi(t)$, and so in the Dirac representation of the $\gamma$ matrices where $\gamma^{0} = {\rm diag}(+1,+1,-1,-1)$, it can be checked that the projection conditions requires that $\psi$ have the form:
\be
\psi_+ = a^{-3/2}\sum_{\omega,k\geq 0} \bv e^{-ikt}\left[A_{+}(\omega,k)e^{\omega t} + A_{-}(\omega,k)e^{-\omega t}\right] \\
e^{-ikt}\left[B_{+}(\omega,k)e^{\omega t} + B_{-}(\omega,k)e^{-\omega t}\right] \\
e^{+ikt}\left[C_{+}(\omega,k)e^{\omega t} + C_{-}(\omega,k)e^{-\omega t}\right] \\
e^{+ikt}\left[D_{+}(\omega,k)e^{\omega t} + D_{-}(\omega,k)e^{-\omega t}\right] \ev. 
\label{psiForm}
\ee
where $A_{+}(0,k) = B_{+}(0,k) = C_{-}(0,k)=D_{-}(0,k) = 0$ and $A_{-}(\omega,0) = B_{-}(\omega,0) = C_{+}(\omega,0)=D_{+}(\omega,0) = 0$.

When $V^{\prime}(\Phi) \equiv V^{\prime}_0 = {\rm const}$, solutions to the $\psi$ field equation are proportional to $\exp(\pm i \sqrt{V^{\prime}_0}t)$ and $\sqrt{-\mathsf{D}^2} \rightarrow \sqrt{V^{\prime}_0}$. Using the form of $\psi$ mandated by the projection condition is then straight-forward to check that:
$$
\tilde{\Psi} = \Vert V^{\prime}_0 \Vert \tilde {\Phi}.
$$
and $\tilde{\Phi} \geq 0$.

Combining this with Eq.~\eqref{eq:PsiDot} we have that $V^{\prime}_0 > 0$ implies $\dot{\tilde{\Phi}} = 0$. If $V^{\prime} < 0$, however, there is no additional restriction on $\dot{\tilde{\Phi}}$. 

The Friedman equation with additional source $\rho_{\rm matter}$ now reads:
\be
H^2 = \frac{\kappa}{3}\left[ \Psi + V(\Phi)+\rho_{\rm matter}\right].
\ee
So the energy density $\rho_{\psi}$ and pressure $p_{\psi}$ are given by:
\be
\rho_{\psi} = \Psi + V(\Phi), \label{eq:rhopsi} \\
p_{\psi} = V^{\prime}(\Phi)\Phi - V(\Phi). \label{eq:ppsi}
\ee
When $V(\Phi) = m^2 \Phi = m^2 \bar{\psi}\psi$ we therefore have $p_{\psi}=0$ and the non-standard spinors (NSS) evolve like dust $\rho_{\psi} \propto a^{-3}$. In this situation, $V^{\prime} = m^2 = {\rm const}$ and so $\Psi = \vert m^2 \vert \Phi \propto a^{-3}$ and so if $m^2 >0$, $\rho_{\psi} = 2 m^2 \Phi \propto a^{-3} \geq 0$ whereas if $m^2 < 0$, $\rho_{\psi} = 0 = p_{\psi}$. The projection condition has essentially ensured the positivity of the potential energy, that is $\rho_{\psi} \geq 0$.  

\subsection{Non-trivial de Sitter Solutions}

We now consider what is required for there to be a de-Sitter solution where $p_{\psi} = -\rho_{\psi}$; $\rho_{\psi} \neq 0$. We distinguish between trivial de-Sitter solutions where the effective dark energy density is $\rho_{\rm de} =V(0)$ and non-trivial ones where $\rho_{\rm de} > V(0)$. All non-trivial solutions feature a cosmological spinor condensate $\Phi = \Phi_0 > 0$, see also~\cite{Gredat:2008qf,Shankaranarayanan:2009sz,Shankaranarayanan:2010st}.

Eqs.~\eqref{eq:rhopsi} and \eqref{eq:ppsi} give that $p_{\psi} = -\rho_{\psi}$ implies $\tilde{\Psi} = -V^{\prime}(\Phi)\tilde{\Phi}$. Differentiating the relationship and using Eq.~\eqref{eq:PsiDot} then gives $V^{\prime}(\Phi) = {\rm const}$. There are then two possibilities. If $V^{\prime \prime} = 0$, so that $V(\Phi) = V_0+ V^{\prime}_0 \Phi$, we have $p_{\psi} = -V_0$ and so all de Sitter solutions must have $\rho_{\psi} = V_0$ i.e. a trivial de-Sitter solution where, other than the contribution from the constant term in $V$, the NSS energy vanishes. 

Non-trivial solutions therefore require $V^{\prime\prime} \neq 0$ and $\Phi=a^{-3}\tilde{\Phi} = \Phi_0 = {\rm const}$ so that $V^{\prime}(\Phi) = {\rm const}$. de-Sitter solutions have $H = \dot{a}/a = H_0 = {\rm const} > 0$ and so $\tilde{\Phi} = \Phi_0 e^{3H_0 t}$.  Using the Friedman equation, $H^2=H^2_0 = \kappa \rho_{\psi}/3$ and Eq.~\eqref{eq:PhiDDot} then gives:
\be
H^2_0 = -\frac{4}{9}V^{\prime}(\Phi_0) = \frac{\kappa}{3}\left[ V(\Phi_0)-V^{\prime}(\Phi_0)\Phi_0\right].
\ee
Since $H^2_0 > 0$, we must have $V^{\prime}(\Phi_0)<0$. We also need:
\be
V^{\prime}(\Phi_0) = -\frac{V(\Phi_0)}{\frac{4}{3\kappa}-\Phi_0}. \label{Phi0Eqn}
\ee
If $V(\Phi_0) > 0$ then we need $\Phi_0 < 4/3\kappa$.

In this non-trivial de Sitter, $\psi = \psi_0$. Noting that $a^{-3/2} = e^{-3H_0 t/2}$, we see from Eq.~\eqref{psiForm} that the projection condition requires the following form (in the Dirac representation) for $\psi_0$:
\be
\psi_0 = A_0 \bv \cos \theta \\ e^{i\phi}\sin \theta \\ 0 \\ 0 \ev.
\ee
for some $\theta \in [-\pi/2, \pi/2]$, $\phi \in [0, 2\pi)$ and non-vanishing $A_0 \in \mathbb{C}$. We then have $\Phi_0 = \Vert A_0\Vert^2 > 0$.

We note, in passing, that such de Sitter solutions exist for quartic potentials (in $\psi$) i.e.
$$ 
V(\bar{\psi}\psi) = V(\Phi) = \lambda\left[\mu^4 + \frac{1}{2}\left(\Phi - m^2\right)^2\right],
$$
for $\lambda >0$. It is clear that $V(\Phi)$ is non-negative for all real values the $\mu$ and $m^2$ and solving the equation for $\Phi_0$ requiring $V^{\prime}(\Phi) < 0$ gives:
\be
\Phi_0 = \frac{4}{3\kappa} - \sqrt{2\mu^4 + \left(\frac{4}{3\kappa}-m^2\right)^2}
\ee
The projection condition on $\psi_0$ requires that $\Phi_0 > 0$ and this requires:
\be
\frac{4 m^2}{3\kappa} > \frac{1}{2}m^4 + \mu^4,
\ee
which is certainly satisfied when $m, \mu^2/m \ll M_{\rm pl}$. With such a potential the effective dark energy density is $\rho_{\psi}=\rho_{\rm de} = \lambda(\mu^4 + m^4/2 - \Phi_0^2/2)$.

\subsection{Perturbations about the de Sitter Solution and Stability}

To consider the stability of NSS de-Sitter solutions, we rewrite Eqs.~\eqref{eq:PhiDDot} and \eqref{eq:PsiDot} in terms of $\Phi = a^{-3}\tilde{\Phi}$ and $\Psi = a^{-3}\tilde{\Psi}$. We have:
\be
\Phi_{pp} &+& \left[6-F\right]\Phi_{p} = \left[\frac{4}{\kappa}-3\Phi\right] F -3\Phi G, \nonumber \\
\Psi_p &+& 3\Psi = -V^{\prime}(\Phi)\left[\Phi_p + 3\Phi\right], \\
F &=& -\frac{\dot{H}}{H^2} = \frac{3}{2}\frac{\Psi + V^{\prime}(\Phi)\Phi}{ \Psi + V(\Phi)}, \\
G &=& \frac{\left(\frac{4}{\kappa} -3\Phi\right) V^{\prime}(\Phi) + 3 V(\Phi)}{\Psi+V(\Phi)},
\ee
where $p = \ln a$. In the de-Sitter background $F=G=0$ and $\Phi = \Phi_0 = {\rm const}$. We now consider a linear perturbation, $\delta \Phi$, in $\Phi$. Now:
\be
3\delta G &=& \frac{4V^{\prime \prime}(\Phi_0)}{H_0^2} \left[1-\frac{3\kappa \Phi_0}{4}\right] \delta \Phi, \nonumber\\
\frac{4}{\kappa} \delta F &=& \frac{2 \left(\delta \Psi + V^{\prime}(\Phi_0)\delta \Phi\right)}{H_0^2} + \frac{2 V^{\prime\prime}(\Phi_0)\Phi_0}{H_0^2}\delta \Phi. \nonumber
\ee
We define $\delta Y = (\delta \Psi + V^{\prime}(\Phi_0)\delta \Phi)/H_0^2$ and let:
\be
\epsilon_0 = 1-\frac{3\kappa \Phi_0}{4}, \qquad \eta_0 = V^{\prime \prime}(\Phi_0)\Phi_0/H^2_0. \nonumber
\ee
We then have:
\be
\delta \Phi_{pp} &+& 6\delta \Phi_{p} = 2\epsilon_{0}\left[ \delta Y - \eta_0 \delta \Phi\right], \nonumber \\
\delta Y_{p} &+& 3\delta Y= -3\eta_0 \delta \Phi. \nonumber
\ee
Now defining $\delta \tilde{\Phi} = a^3 \delta \Phi$ and $\delta \tilde{Y} = a^3 \delta Y$ we and differentiating the $\delta \Phi$ equation we have:
\be
\delta \tilde{\Phi}_{ppp} - 9\delta\tilde{\Phi}_{p} = -2\epsilon_{0}\eta_0\left[3\delta \tilde{\Phi} + \delta \tilde{\Phi}_{p}\right].
\ee
Thus $\delta \Phi = \Phi_0 \sum_{q} Q_{q} e^{H_0 q t}$ for some constants $Q_{q}$ where:
$$
\left[q(3+q) + 2\epsilon_0 \eta_0\right](q+6) = 0.
$$
So either $q = -6$, or $q = q_{\pm}(\epsilon_0 \eta_0)$ where:
\be
q_{\pm}(\epsilon_0 \eta_0) = \frac{3}{2}\left[-1 \pm \sqrt{1-\frac{8 \epsilon_0 \eta_0}{9}}\right]. \label{qEqn}
\ee
Since the underlying field $\psi$ obeys a second order equation and has a projection condition to eliminate ghost modes, it may seem slightly strange that there are three linearly independent modes in $\delta \Phi$ (and hence $\delta \rho_{\psi}$). Fortunately, as we show below, the projection condition ensures that the $q_{-}$ is not actually present ($Q_{q_{-}} = 0$).

Note that perturbations of the spinorial part of a spinor in this context have been considered in~\cite{Gredat:2008qf,Shankaranarayanan:2009sz,Shankaranarayanan:2010st} where a hedgehog type ansatz was used to identify the correct degrees of freedom when perturbing a spinor in a cosmological spacetime.

Consider the form of $\psi$. We have $\tilde{\psi} = a^{-3/2}\psi = a^{-3/2}\left[e^{3H_0t/2} \psi_0 + \delta \tilde{\psi}\right]$. We then have
\be
\partial^2_{t} \delta \tilde{\psi} = -V^{\prime}(\Phi_0)\delta \tilde{\psi} - e^{3H_0t/2} V^{\prime \prime}(\Phi_0)\psi_{0} \delta \Phi + \delta\tilde{\chi}_{-}.
\ee
where $\tilde{P}\delta\tilde{\chi}_{-} = -\delta\tilde{\chi}$. Solving this equation with the projection condition it follows that:
\be
\delta \tilde{\psi} = - \frac{Q_{q_{+}}}{2 \epsilon_{0}} e^{(\frac{3}{2}+q)H_0 t} \psi_0 + C_0 e^{\frac{3}{2}H_0 t} \psi_0 + C_1 e^{\frac{3}{2}H_0 t}\psi_1 - C_2 e^{-\frac{3}{2} H_0 t} \psi_2,
\ee
where $C_i \in \mathbb{C}$ and
$$
\psi_1 = A_0 \bv \sin \theta \\ -e^{i\phi}\cos \theta \\ 0 \\ 0 \ev, \qquad \psi_2 = A_0 \bv 0 \\ 0 \\ \cos \beta \\ e^{i\alpha}\sin \beta \ev,
$$
for some $\alpha$ and $\beta$. We note that $\bar{\psi}_{1}\psi_1 = \Phi_0$, $\bar{\psi}_{2}\psi_2 = -\Phi_0$ and that both $\psi_{1}$ and $\psi_{2}$ are orthogonal to $\psi_0$. Thus writing $a = e^{H_0 t + \delta A}$ and $Q_{q_{\pm}}= Q_{\pm}$ we have:
\be
\frac{\Phi}{\Phi_0} = 1 -3\delta A + \frac{Q_{+}}{\epsilon_0}e^{H_0 q_{+} t} + 2{\rm Re}(C_0)+ \vert C_1\vert^2 - \vert C_2 \vert^2 a^{-6}.
\ee
We then have:
\be
3\delta A_{p} = \frac{\kappa}{2H_0^2} \left[\delta \Psi + V^{\prime}(\Phi_0)\delta \Phi\right].
\ee
It follows that:
\be
-3\delta A &=& -3A_1 - \frac{(\epsilon_0 -1)}{\epsilon_0} Q_{+} e^{H_0 q_{+} t} \nonumber \\ &&- \frac{(\epsilon_0 -1)}{\epsilon_0} Q_{-} e^{H_0 q_{+} t} + \frac{(\epsilon_0-1)\mu_0}{9}Q_{-6}e^{-6 H_0 t}. \nonumber
\ee
Thus:
\be
\frac{\Phi}{\Phi_0} &=& 1 + Q_{+} a^{q_{+}} + \left[(2{\rm Re}(C_0) + \vert C_1\vert^2 - 3A_1\right] \\
&& - \frac{(\epsilon_0 -1)}{\epsilon_0} Q_{-} a^{q_-} \nonumber +\left(\frac{(\epsilon_0-1)\mu_0}{9}Q_{-6}- \vert C_2\vert^2\right)a^{-6}. \nonumber
\ee
Now we previously found that $\Phi/\Phi_0 = 1 + Q_{+}a^{q_{+}} + Q_{-}a^{q_{-}} + Q_{-6}a^{-6}$ and so since $1-\epsilon_0 = 3\kappa \Phi_0/4 > 0$ we must have $Q_{-} = 0$ and:
$$
Q_{-6}\left[1+\frac{\kappa \Phi_0 \mu_0}{12}\right] = -\vert C_2\vert^2.
$$
Thus we see that the projection condition eliminates one of the possible modes in $\delta \Phi$ (that proportional to $a^{q_{-}}$ and we have:
$$
\delta \Phi = \Phi_0 \left[ Q_{+}a^{q_{+}} - \frac{\vert C_2\vert^2 a^{-6}}{1+\frac{\kappa \Phi_0 \mu_0}{12}}\right],
$$
for some $Q_{+}$ and $C_{2}$. Finally, we that $\delta \rho_{\psi} = \delta \Phi + V^{\prime}(\Phi_0)\delta \Phi$ is given by:
\be
\delta \rho_{\psi} = \frac{Q_{+} a^{q_{+}}}{3+q_{+}} + \frac{\vert C_2\vert^2 a^{-6}}{3+\frac{\kappa \Phi_0 \mu_0}{4}}.
\ee
If $\mu_0 \epsilon_0 > 9/8$ then we must take $q_{+} = -3/2 - i \sqrt{8\mu_0\epsilon_0/9-1}$ in the expression for $\psi$ and replace $a^{q_{+}}$ with ${\rm Re}(a^{q_{+}})$ in $\delta \Phi$. In $\delta \rho_{\psi}$ we take the real part of $a^{q_{+}}/(q_{+}+3)$.

\section{Conclusions}
\label{sec:conc}

In this article, we have constructed a new class of theories of non-standard spinors (NSS). Their dynamics is more general that than that of Dirac or Majorana spinors, even when self-interactions are taken into account. In contrast to standard spinors, the dynamics of NSS is \emph{not} described by a first order equations of motion like the Dirac equation. This leads to a more general and thus more interesting cosmological behavior than that exhibited by normal spinors, including for instance the existence of non-trivial de Sitter solutions. It is therefore possible to invoke NSS, as an alternative to scalar fields, as one possible explanation of the early and late time acceleration of our Universe. As example of a NSS theory is that of the eigenspinors of $C$, originally proposed by Ahluwalia-Khalilova and Grumiller in Ref.~\cite{jcap,prd}. 

We have constructed a general action for NSS. We began by considering a Klein-Gordon action for spinors and noted that because there is no positive definite Lorentz invariant norm for spinors, such an action would have propagating negative energy ghost modes. These would lead to instabilities in the quantum theory. These negative energy modes can, however, be eliminated by including an additional term in the action. This term depends on an operator $\opp{P}$, which must have the property that $\opp{P} \psi = \psi$ on positive energy modes, and $\opp{P}^2 \psi = -\psi$ on negative energy ones. Hence $\opp{P}^2 = \mathbb{I}$. We also noted that in momentum space $P$ must be an odd function of momentum i.e. $\opp{P}(\mathbf{p}) = -\opp{P}(-\mathbf{p})$.  The original ELKO model as well as Dirac and Majorana spinors correspond simply to specific choices of the projection operator $\opp{P}$. 

By constructing the NSS action in this way, we found that ELKO spinors require a choice of $\opp{P}$ that is \emph{not} Lorentz invariant but instead includes a preferred axis. Previous works on this field have effectively made a specific choice of frame so that this preferred direction and hence the violation of Lorentz invariance was not manifest explicitly. Using our general definition of NSS theories, however, the violation of Lorentz invariance which is required to define ELKOs is clear at the level of the action.  We note that an alternative definition of the eigenspinors of $\opp{C}$ replaces the spatial direction with a preferred time-like direction.

A truly Lorentz invariant NSS theory requires a Lorentz invariant projection operator $\opp{P}$. For most such choices of the operator, $\opp{P} \psi = \psi$ is essentially equivalent to the Dirac equation with self-interaction terms. The projection condition then effectively reduces the dynamics from second to first order equations of motion. We found that there was only one suitable, Lorentz invariant choice of $P$ which preserves the second order dynamics. In momentum space and by assuming a flat background this operator is given by $\opp{P} = p_{\mu}\gamma^{\mu}/\sqrt{p_{\mu}p^{\mu}}$, and so $\opp{P}$ is a non-local operator. In the absence of self-interactions, i.e.~we have $V = V_0 + m^2 \bar{\psi}\psi$, the field equations then reduce to the Dirac equation, but for more complicated choices of $V$ this is no longer the case.

Having provided a general definition of NSS we then constructed and examined the full energy-momentum tensor. In the case of ELKOs we noted that even if one ignores the additional contributions to the energy-momentum tensor from the variation of $P$ which respect to the metric, the energy momentum tensor differs from that which has previously appeared in the literature~\cite{Boehmer:2006qq,Boehmer:2007dh,n6,Boehmer:2007ut,Boehmer:2008ah,Boehmer:2008rz,Gredat:2008qf,Boehmer:2009aw,n1,n3,n4,n5,Shankaranarayanan:2009sz,Shankaranarayanan:2010st,Boehmer:2010tv,Wei:2010ad}. We show explicitly in Appendix~\ref{appelko} where the additional terms come from, namely from the variation of the spin connection with respect to the metric. In case of Dirac spinors this contribution identically vanishes, see Appendix~\ref{appdirac}, and therefore, we believe, this has been neglected in the past for ELKO spinors.

The presence of additional terms even in the ELKO energy-momentum tensor led us to re-address the cosmology of such models. We defined $\psi = \varphi(t)\xi$, where $\xi$ is a constant spinor, and so were able to treat $\varphi$ as the only dynamical variable cosmological. In the simplest case, the flat FLRW background, it produces an effective gravitational coupling $G$ which places a simple limit on the maximum value of $\Phi$, namely $\Phi < 1 / \sqrt{\pi G} = 2 \sqrt{2}M_{Pl}$. When we examine de Sitter type solutions, this ansatz produces a potential very similar to those discussed in previous work. Its form is surprisingly similar to the case where the variation of the spin connection with respect to the metric was neglected. However, as we noted, it is not clear if a Lorentz invariant NSS model can be found which has these dynamics.

We also considered the dynamics of the only Lorentz invariant NSS model we were able to construct. We found that cosmologically the dynamics can be written in terms of $\Phi = \bar{\psi}\psi$ and $\Psi = \dot{\bar{\psi}}\dot{\psi}$. The potential describing self-interactions depends only on $\Phi$: $V=V(\Phi)$. When $V_{,\Phi\Phi} =0$, we had previously noted that the theory should be equivalent to that of a Dirac spinor, and so $\rho_{\psi}\propto a^{-3}$. We confirmed that this was indeed the case. When $V_{,\Phi \Phi} \neq 0$ and $V_{,\Phi} < 0$ we found that there stable de Sitter solutions; stability of these solutions is ensured when $V_{,\Phi \Phi} > 0$. In contrast to the situation with scalar fields, de Sitter solutions do \emph{not} required $\vert V_{,\Phi}/\kappa V(\Phi)\vert \ll 1$ (i.e. slow-roll). Instead with NSS we generally have $-V_{,\Phi}/\kappa V(\Phi) \sim O(1)$. With NSS spinors, the expansion of the Universe acts as a brake to prevent the effective scalar field $\Phi$ rolling down the potential.

The main results of the paper lie in our discussion of the definition and dynamics of the entire class of NSS, and their cosmology. We laid the foundations of an in depth analysis of the dynamics of this field in an arbitrary spacetime with a focus on cosmological dynamics. Importantly we also constructed what is, to the best of our knowledge, the only Lorentz invariant, ghost free proposal for a theory of non-standard spinors.

The cosmological dynamics of the effective scalar degree of freedom in both ELKO and Lorentz invariant NSS cosmology show a large number of very interesting properties, mainly due to their more complicated couplings to the gravitational sector when compared to the scalar field. The cosmological evolution of the NSS energy density exhibits a much wider range of behavior than that seen with Dirac spinors where it always scales as $a^{-3}$. The existence of stable de Sitter solutions means that NSS could represent an alternative to scalar field inflation / dark energy. 

\begin{acknowledgments}
JB would like to thank Dharam Ahluwalia for discussions about this manuscript and for kind hospitality during the 1st International Workshop On Elko, 26 Feb -- 5 Mar 2010, Christchurch, New Zealand. JB further thanks Adam Gillard, Sebastian Horvath, Cheng-Yang Lee, Ben Martin and Dimitri Schritt. DFM thanks the Research Council of Norway for the FRINAT grant 197251/V30. DFM is also partially supported by project CERN/FP/109381/2009 and PTDC/FIS/102742/2008.
\end{acknowledgments}

\bibliographystyle{JHEP}
\bibliography{review}

\appendix

\section{Spin Connection Contribution to the Energy-Momentum Tensor}
The purpose of these appendices is to demonstrate that the contribution from the variation of the spin connection, $\omega_{\mu}^{ab}$ to the energy momentum tensor has generally be overlooked in previous work on ELKO spinors. First we look in detail at the derivation of $\delta \Gamma_\mu$ then we will check with the Dirac spinor that this does not give any contribution, as expected, but rather in the case of an ELKO spinor it becomes very important. 

\subsection{Variation of $\delta\Gamma_\mu$}

Starting from \eqref{eq:spincon} we can write down the variation
\be
\delta \Gamma_{\mu} &=& \frac{i}{4}f_{ab} \delta \omega_{\mu}^{ab}.
\ee
Now
\be
\delta \omega_{\mu}^{ab} = \delta e_{\nu}^{a} \nabla_{\mu} e^{\nu b} + e_{\nu}^{a}\nabla_{\mu} \delta e^{\nu b} + e_{\nu}^{a}e^{\sigma b}\delta \Gamma_{\mu \sigma}^{\nu}.
\ee
We evaluate this in a local inertial frame where $g_{\mu \nu,\rho} =0$ and $e_{\mu,\rho}^{a} =0$, in such a LIF:
\be
\left.\delta \omega_{\mu}^{ab}\right\vert_{\rm LIF} &=& -  e^{\nu b} \partial_{\mu} \delta e_{\nu}^{a} + \frac{1}{2}e^{\nu a} e^{\sigma b}\left[\partial_{\sigma}\delta g_{\mu \nu} + \partial_{\mu} \delta g_{\nu \sigma} - \partial_{\nu} \delta g_{\mu \sigma}\right], \\
&=&e^{\nu [a}e^{\sigma b]}\partial_{\mu}\left[e_{\nu c}\delta e_{\sigma}{}^{c}\right] + e^{\nu [a}e^{\sigma b]} \partial_{\sigma} \delta g_{\mu \nu}. \nonumber
\ee
Hence in a general frame:
\be
\delta \Gamma_{\mu} = \frac{i}{4}f^{\nu \rho} \left[ \nabla_{\mu}\left(e_{[\nu c}\delta e_{\rho]}{}^{c}\right) + \nabla_{\rho} \delta g_{\mu \nu}\right] \nonumber
\ee
where $f^{\nu \sigma} = e^{\nu}_{a}e^{\sigma}_{b}f^{ab}$.  Now $\delta g_{\mu \nu} = 2 e_{(\mu}^{c}\delta e_{\nu) c}$ and so the first term on the right hand side does not contribute to the variation with respect to $g_{\mu\nu}$, and so just varying $g_{\mu \nu}$ we have:
\be
\delta \Gamma_{\mu} = \frac{i}{4}f^{\nu \rho}\nabla_{\rho}\delta g_{\mu \nu} = -\frac{i}{4}f^{\rho \nu}\nabla_{\rho}\delta g_{\mu \nu}.
\ee

\subsection{Dirac Spinors}
\label{appdirac}

In this subsection we will calculate the contribution to the Dirac energy-momentum tensor from $\delta \Gamma_\mu$ which will be zero. We follow Hehl~\cite{Hehl:1973,Hehl:1974,Hehl:1976kj} and write the Lagrangian as
\be
\mathcal{L}_{\rm D} = \frac{1}{2}\left(\bar\psi i \gamma^\alpha \nabla_\alpha \psi -  \bar\psi \overleftarrow{\nabla}_\alpha i \gamma^\alpha \psi\right) - m \bar{\psi}\psi.
\ee

Thus we find
\be
T^{\mu \nu}_{\rm D} &=& -\frac{2}{\sqrt{-g}}\frac{\delta (\sqrt{-g}\mathcal{L}_{\rm D})}{\delta g_{\mu \nu}}  \\ &=&\frac{i}{2}\left[ \bar{\psi} \gamma^{(\mu} \nabla_{\nu)}\psi - \bar{\psi}\overleftarrow{\nabla}^{(\mu} \gamma^{\nu)} \psi \right] - g^{\mu \nu} \mathcal{L}_{\rm D} + T^{\prime \mu \nu}, 
\ee
where $T^{\prime \mu \nu}$ includes just the contributions from $\delta \Gamma_{\alpha}$.  We will confirm that in the case of Dirac spinors, $T^{\prime \mu \nu}$ vanishes.

Thus:
\be
T^{\prime \mu \nu} = i \left[  \bar{\psi} \gamma^{\alpha} \frac{\delta \Gamma_{\alpha}}{\delta g_{\mu \nu}} \psi + \bar{\psi} \delta{\Gamma_{\alpha}}{\delta g_{\mu \nu}}\gamma^{\alpha} \psi\right],
= \frac{1}{4} \nabla_{\rho}\left[ \bar{\psi} F^{(\mu \nu) \rho} \psi\right], \nonumber
\ee
where
\be
F^{\mu \nu \rho} = \gamma^{\mu} f^{\nu \rho} + f^{\nu \rho}\gamma^{\mu} = \frac{i}{2}\left[\gamma^{\mu}\gamma^{\nu}\gamma^{\rho} - \gamma^{\rho}\gamma^{\nu}\gamma^{\mu}  - \gamma^{\mu} \gamma^{\rho}\gamma^{\nu} + \gamma^{\nu}\gamma^{\rho} \gamma^{\mu}\right] = -F^{\nu \mu \rho},
\ee
where we have used $\gamma^{\mu}\gamma^{\nu} = -\gamma^{\nu}\gamma^{\mu} + 2g^{\mu \nu}$ to establish the antisymmetry in $\mu$ and $\nu$.  It follows that $F^{(\mu \nu)\rho} \equiv 0$ and hence $T^{\prime \mu \nu} \equiv 0$ also.  As expected, the contribution to $T^{\mu \nu}$ from varying the spin connection is identically zero in the case of Dirac spinors.

\subsection{NSS}
\label{appelko}

In this subsection we consider the contribution to the energy momentum tensor from the NSS action when the operator $P$ is fixed i.e. $\delta P =0$. We show that ignoring the contribution from $\delta \Gamma_\mu$ gives an incomplete energy-momentum tensor.

We focus on the variation of $\mathcal{L}_{\psi}^{(2)}$ with respect to the metric; the variation of $\mathcal{L}_{\psi}^{(1)}$ is similar.  Fixing $P$, modulo the projection condition, the Lagrangian is
\be
\mathcal{L}^{(2)}_{\psi} = g^{\mu \nu}  \dual{\psi}\overleftarrow{\nabla}_{\mu} \nabla_\nu \psi
\ee
When we vary this with respect to the metric we find:
\be
T^{(2) \mu \nu}_{\psi} = -\frac{2}{\sqrt{-g}}\frac{\delta (\sqrt{-g}\mathcal{L}^{(2)}_{\psi})}{\delta g_{\mu \nu}} = \left[2\dual\psi \overleftarrow{\nabla}^{(\mu} \nabla^{\nu)} \psi - \mathcal{L}^{(2)}_{\psi} g^{\mu \nu}\right] + T^{\prime \mu \nu} ,
\ee 
where
\be
\label{eq:fullvar}
T^{\prime \mu \nu} &=&  -2g^{\rho \sigma} ( \dual{\psi} \frac{\delta\Gamma_{\rho}}{\delta g_{\mu \nu}} \nabla_\sigma \psi -
 \dual{\psi} \overleftarrow{\nabla}_{\rho} \frac{\delta \Gamma_{\sigma}}{\delta g_{\mu \nu}} \psi )\nonumber \\
&=& \frac{i}{2}g^{\rho \sigma}\nabla_{\kappa}\left[ \dual{\psi} f^{\kappa \mu}\delta_{\rho}^{\nu} \nabla_{\sigma}\psi - \dual{\psi} \overleftarrow{\nabla}_{\rho} f^{\kappa \mu}\delta_{\sigma}^{\nu} \psi\right], \nonumber \\
&=&  \nabla_{\rho}J^{\mu \nu \rho}, \nonumber
\ee
where we have defined:
\be
J^{\mu \nu \rho} = -\frac{i}{2}\left[  \dual{\psi} \overleftarrow{\nabla}^{(\mu}f^{\nu)\rho} \psi +\dual{\psi} f^{\rho (\mu} \nabla^{\nu)}\psi\right].
\ee
Thus the full energy-momentum term (up to an additional contribution from the variation of $P$) is:
\be
T_{\mu \nu}^{(2)\psi} &= & 2\nabla_{( \mu} \dual{\psi} \nabla_{\nu )}\psi - \frac12 g_{\mu \nu}\mathcal{L}  \\ &&+ \nabla_{\rho}J_{\mu \nu}{}^{\rho}. \nonumber
\ee
The first line is the usual energy-momentum of the Elko spinor quoted in literature~\cite{Boehmer:2006qq,Boehmer:2007dh,Boehmer:2007ut,Boehmer:2008ah,Boehmer:2008rz}. The second line provides the additional contributions to $T_{\mu \nu}$ from variation of the metric in the spin connection. The important point is that this is non zero. 

\section{Path Integral Quantization}\label{app:quant}
In this appendix, we consider the quantization of the action:
\be
S[\psi,\dual{\psi},\chi,\dual{\chi}] &=& \int \sqrt{-g}\, \left(\mathcal{L}_{\psi}^{(1)}+\mathcal{L}_{P}\right)\dd^4x, \\
\mathcal{L}_{\psi}^{(1)} &=& (\dual{\psi} \overleftarrow{\slashed{\nabla}})(\slashed{\nabla}\psi) -V(\dual{\psi},\psi), \\
\mathcal{L}_{P} &=& -\dual{\chi}\opp{P}_{-}\psi - \dual{\psi}\oppL{P}_{-}^{(A)}\chi.
\ee
using the path integral formalism.  We define $V(\dual{\psi},\psi) = m^2 \dual{\psi}\psi + I(\dual{\psi},\psi)$ so that $I(\dual{\psi},\psi)$ represents the non-free field self interaction terms.  We will assume that the effect of $I(\dual{\psi}.\psi)$ can be evaluated via a perturbative expansion about the free field theory. We can split $S$ into a free-field action $S_0$ and interaction term $I_{\psi}[\dual{\psi},\psi]$ thus:
\be
S[\psi,\dual{\psi},\chi,\dual{\chi}] = S_{0}[\psi,\dual{\psi},\chi,\dual{\chi}] - I[\dual{\psi},\psi].  \label{Saction}
\ee
where the free field action is:
\be
S_0[\psi,\dual{\psi},\chi,\dual{\chi}] = \int \sqrt{-g}\, \left( (\dual{\psi} \overleftarrow{\slashed{\nabla}})(\slashed{\nabla}\psi) -m^2 \dual{\psi}\psi+\mathcal{L}_{P}\right)\dd^4x. \nonumber
\ee
The interaction term is therefore given by:
\be
I_{\psi}[\dual{\psi},\psi] = \int \sqrt{-g} \left[V(\dual{\psi},\psi) - m^2 \dual{\psi}\psi\right]\dd^4 x. \nonumber
\ee
Such an interaction term is generally introduced perturbatively and so the free field mass, $m^2$, should be chosen so that $I_{\psi}$ can be treated as a small perturbation to $S_0$.  

The generating functional, $Z[J,\dual{J}]$ is defined by:
\be
Z[J,\dual{J}\,] &=& \int\, \mathcal{D}\psi\mathcal{D}\dual{\psi}\mathcal{D}\chi\mathcal{D}\dual{\chi}\, e^{iS[\psi,\dual{\psi},\chi,\dual{\chi}] + i\int \sqrt{-g}\left(\dual{J}\psi +\dual{\psi}J\right)\dd^4 x}.
\ee
where $J$ and $\dual{J}$ are respectively spinor and dual spinor valued current. Quantum expectations of the form $\left\langle A(\dual{\psi},\psi) \right\rangle$, for some function $A$ are the given by:
\be
\left\langle A(\dual{\psi},\psi) \right\rangle &=& \frac{\int\, \mathcal{D}\psi\mathcal{D}\dual{\psi}\mathcal{D}\chi\mathcal{D}\dual{\chi}\, A(\dual{\psi},\psi) e^{iS[\psi,\dual{\psi},\chi,\dual{\chi}]}}{\int\, \mathcal{D}\psi\mathcal{D}\dual{\psi}\mathcal{D}\chi\mathcal{D}\dual{\chi}\, e^{iS[\psi,\dual{\psi},\chi,\dual{\chi}]}}, \nonumber \\
&=&  \left[ A\left(\frac{\delta}{i\sqrt{-g}\delta J}, \frac{\delta}{i\sqrt{-g}\delta \dual{J}}\right) \ln Z[J,\dual{J}\,]\right]_{J=\dual{J}=0}. \nonumber
\ee
It is clear that observables are unaltered by a ($J$, $\dual{J}$)-independent rescaling of $Z[J,\dual{J}\,]$.

Using Eq. (\ref{Saction}) we may rewrite $Z[J,\dual{J}\,]$ thus:
\be
Z[J,\dual{J}\,] &=& \int\, \mathcal{D}\psi\mathcal{D}\dual{\psi}\mathcal{D}\chi\mathcal{D}\dual{\chi}\, \exp\left(-i I_{\psi}[\dual{\psi},\psi] \right)e^{iS_{0}[\psi,\dual{\psi},\chi,\dual{\chi}] + i\int \sqrt{-g}\left(\dual{J}\psi +\dual{\psi}J\right)\dd^4 x} \nonumber \\
&=& \exp\left(-i I_{\psi}\left[\frac{\delta}{i\sqrt{-g}\delta J},\frac{\delta}{i\sqrt{-g}\delta \dual{J}} \right] \right) \left[\int\, \mathcal{D}\psi\mathcal{D}\dual{\psi}\mathcal{D}\chi\mathcal{D}\dual{\chi}\, e^{iS_{0}[\psi,\dual{\psi},\chi,\dual{\chi}] + i\int \sqrt{-g}\left(\dual{J}\psi +\dual{\psi}J\right)\dd^4 x}\right] \nonumber  \\
&=& \exp\left(-i I_{\psi}\left[\frac{\delta}{i\sqrt{-g}\delta J},\frac{\delta}{i\sqrt{-g}\delta \dual{J}} \right] \right) Z_{0}[J,\dual{J}\,], \label{Zeqn}
\ee
where $Z_0$ is the free field generating functional given by:
\be
Z_0[J,\dual{J}\,] = \int\, \mathcal{D}\psi\mathcal{D}\dual{\psi}\mathcal{D}\chi\mathcal{D}\dual{\chi}\, e^{iS_{0}[\psi,\dual{\psi},\chi,\dual{\chi}] + i\int \sqrt{-g}\left(\dual{J}\psi +\dual{\psi}J\right)\dd^4 x}.
\ee

\subsection{Free Field Propagator}\label{app:quant:free}
In flat-space (in coordinates where $\sqrt{-g} = 1$), the field free Green's function, $G_{F}(x-y;m)$, and its Fourier transform, the free-field propagator, $G_F(p;m)$ are 
derived from $Z_0[J,\dual{J}\,]$ by:
\be
Z_{0}[J,\dual{J}\,] &=& Z_{0}[0,0]e^{-i\int \dual{J}(x) G_{F}(x-y;m)J(y) \dd^4 x \dd^4 y}. \label{GFdef}
\ee

Integrating by parts and assuming that all resultant surface terms vanish:
\be
S_{0}[\psi,\dual{\psi},\chi,\dual{\chi}] &+& i\int \sqrt{-g}\left(\dual{J}\psi +\dual{\psi}J\right)\dd^4 x = \int \sqrt{-g}\left\lbrace \dual{\psi}\left[ -\slashed{\nabla}^2 - m^2\right]\psi  \right. \\ \nonumber && \left. - \left(\dual{\chi}\oppL{P}_{-}-\dual{J}\right)\psi -\dual{\psi}\left(\opp{P}_{-}\chi-J\right) \right\rbrace \dd^4 x. \nonumber
\ee
Now define:
\be
\phi &=&\psi - \left[-\slashed{\nabla}^2-m^2\right]^{-1}\left[\opp{P}_{-}\chi-J\right],  \nonumber \\
\dual{\phi} &=&\dual{\psi} - \left[\dual{\chi}\oppL{P}_{-}-\dual{J}\right]\left[-\slashed{\nabla}^2-m^2\right]^{-1}, \nonumber \\
\nu &=& \chi-J, \qquad \dual{\nu} = \dual{\chi}-\dual{J}. \nonumber
\ee
and then:
\be
S_{0}[\psi,\dual{\psi},\chi,\dual{\chi}] &+& i\int \sqrt{-g}\left(\dual{J}\psi +\dual{\psi}J\right)\dd^4 x  = \int \sqrt{-g} \left\lbrace \dual{\phi}\left[ -\slashed{\nabla}^2 - m^2\right]\phi  \right. \\ \nonumber && -  \left.\left[\dual{\nu} \oppL{P}_{-}-\dual{J}\oppL{P}_{+}\right] \left[-\slashed{\nabla}^2-m^2\right]^{-1}\left[\opp{P}_{-}\nu-\opp{P}_{+}J\right] \right\rbrace \dd^4 x. \nonumber
\ee
To quantize this theory in the usual way we must ensure that the terms in the above expression which involve mixing of the currents, $J$ and $\dual{J}$, and the auxiliary fields, $\nu$ and $\bar{\nu}$, vanish.  This requires that  $\opp{P}_{-}\left[\nabla^2 + m^2\right]\opp{P}_{+} = \opp{P}_{+}\left[\nabla^2 + m^2\right]\opp{P}_{-}=0$ when acting on any spinor state.  This is equivalent to requiring that $P$ is chosen so that it commutes with $\slashed{\nabla}^2$:
\be
\left[\opp{P},\slashed{\nabla}^2\right] = 0. \label{eq:Pcommute}
\ee
This is certainly the case with the Lorentz invariant choice $\opp{P}$ given in \S \ref{sec:LoInNSS}.  If we instead consider the quantization of the action with $\mathcal{L}_{\psi}^{(1)} \rightarrow \mathcal{L}_{\psi}^{(2)}$ we would have $\slashed{\nabla}^2 \rightarrow \nabla^2$ everywhere and instead require the $\opp{P}$ commute with $\nabla^2$.  We have not found any choice of $\opp{P}$ with this property in a general background.  Thus with the Lorentz invariant choice of $\opp{P}$ the kinetic structure of $\mathcal{L}_{\psi}^{(1)}$ is clearly preferred.

With $\opp{P}$ obeying Eq. (\ref{eq:Pcommute}), we then have:
\be
S_{0}[\psi,\dual{\psi},\chi,\dual{\chi}] + i\int \sqrt{-g} \left(\dual{J}\psi +\dual{\psi}J\right) \dd^4 x &=& \int \sqrt{-g} \left\lbrace \dual{\phi}\left[ -\slashed{\nabla}^2 - m^2\right]\phi  \right. \nonumber \\ \nonumber && -  \left. \dual{\nu}\oppL{P}_{-}  \left[-\slashed{\nabla}^2-m^2\right]^{-1}\opp{P}_{-}\nu \right. \\ && \left.- \dual{J}\, \oppL{P}_{+}  \left[-\slashed{\nabla}^2-m^2\right]^{-1}\opp{P}_{+}J\right\rbrace \dd^4 x \nonumber.
\ee
Hence:
\be
Z_{0}[J,\dual{J}\,] &=& Z_{0}[0,0]e^{-i\int \sqrt{-g}\dual{J}\,\oppL{P}_{+}\left[-\slashed{\nabla}^2-m^2\right]^{-1}\opp{P}_{+}J \dd^4 x} , \label{Z0eqnJJ} \\
Z_{0}[0,0] &=& \int \mathcal{D}\phi\mathcal{D}\dual{\phi} \mathcal{D}\nu\mathcal{D}\dual{\nu} e^{-i\int \sqrt{-g}\dual{\phi}\left[ \slashed{\nabla}^2 + m^2\right]\phi\dd^4 x + i \int \sqrt{-g}\dual{\nu}\oppL{P}_{-}\left[\slashed{\nabla}^2+m^2\right]^{-1}\opp{P}_{-}\nu\dd^4 x}. \nonumber
\ee
We noted previously that observables do not depend on the overall $\left\lbrace J,\, \dual{J}\right\rbrace$-independent normalization of the generating functional and so are independent of $Z_0[0,0]$.  We therefore drop this overall factor in what follows.
Integrating by parts and using $[\opp{P}, \slashed{\nabla}^2] = 0$ and $\opp{P}_{+}\opp{P}_{+} = \opp{P}_{+}$ (which follows from $\opp{P}^2=\mathbb{I}$), we have:
\be
&&\int \dual{J}\, \oppL{P}_{+}  \left[-\slashed{\nabla}^2-m^2\right]^{-1}\opp{P}_{+}J \sqrt{-g}\dd^4 x \label{JsimpEqn}\\ &&= \int \dual{J}\,  \left[-\slashed{\nabla}^2-m^2\right]^{-1}\opp{P}_{+}J \sqrt{-g}\dd^4 x, \nonumber \\
&&= \iint \left[\dual{J}(x) \left[-\slashed{\nabla}^2_{(x)}-m^2\right]^{-1}\opp{P}_{+}(x) \delta^{(4)}(x-y) J(y)\right] \sqrt{-g(x)}\dd^4 x\sqrt{-g(y)}\dd^4 y.\nonumber
\ee
where $\delta^{(4)}(x-y)$ is the Dirac $\delta$-function. 

Henceforth we work in flat-space and pick coordinates so that $\sqrt{-g} = 1$ and $\nabla_{\mu}=\partial_{\mu}$. We define $\partial_{(x)\mu} = \partial/\partial x^{\mu}$. By comparing Eq. (\ref{GFdef}) and  Eq. (\ref{Z0eqnJJ}) and using Eq. (\ref{JsimpEqn})) we may now read off the free field propagator as:
\be
G_{F}(x-y;m) =\left[-\partial^2_{(x)}-m^2\right]^{-1}\opp{P}_{+}(x) \delta^{(4)}(x-y).
\ee
Using
$$
\delta^{(4)}(x-y) = \frac{1}{(2\pi)^4}\int \dd^4 p\, e^{i p \cdot (x-y)},
$$
and $\opp{P}_{+}(x) e^{i p \cdot (x-y)} = \opp{P}_{+}(p^{\mu}) e^{i p\cdot (x-y)}$ where $\opp{P}_{+}(p^{\mu}) = (\mathbb{I}+\opp{P}(p^{\mu}))/2$ we have:
\be
G_{F}(x-y;m) &=& \frac{1}{(2\pi)^4} \int \dd^4 p\,  \frac{\frac12 (\mathbb{I}+\opp{P}(p^\nu)) }{p^2-m^2},  \\
&=& \frac{1}{(2\pi)^4} \int \dd^4 p\, G_{F}(p;m) e^{ip_{\mu}(x^{\mu}-y^{\mu})}, \nonumber
\ee
where $G_{F}(p;m)$ is, by definition, the free-field propagator. We therefore find:
\be
G_{F}(p;m) = \frac{\frac12(\mathbb{I}+\opp{P}(p^\nu)) }{p^2-m^2}.
\ee
Just as with other quantum propagator, $G_{F}(p;m)$, has simple poles at $p=\pm m$ which are dealt with by making a specific choice of integration contour in $G_F(x-y;m)$ which can be written thus:
\be
G_{F}(x-y;m) &=& \frac{1}{(2\pi)^4} \lim_{\epsilon \rightarrow 0} \int \dd^4 p\, \frac{\opp{P}_{+}(p^\nu) e^{ip_{\mu}(x^{\mu}-y^{\mu})}}{p^2-m^2+i\epsilon}. \nonumber
\ee

The reality condition for $G_{F}(x-y;m)$ is that:
$$ \Im{\rm m} \left[\iint \dual{J}(x)G_{F}(x-y)J(y) \sqrt{-g(x)}\dd^4 x\sqrt{-g(y)}\dd^4 y \right] =0.$$
This is equivalent to $\dual{G}_{F}(x-y) = G_{F}(y-x)$ which in turn is the case if any only if $\oppD{P} = \opp{P}$.  This was precisely one of the conditions that was required in the definition of $\opp{P}$ and so does not represent an additional condition on $\opp{P}$.  We have shown explicitly in \S  \ref{sec:LoInNSS}   that the Lorentz invariant choice of $\opp{P}$ obeys this condition.

In the Lorentz invariant NSS theory we have proposed: 
$$\opp{P}(p^{\mu}) = \frac12(1+i\gamma^5) \opp{P}_{0}(p^{\mu}) +  \frac12(1-i\gamma^{5}) \oppD{P}_0(p^{\mu}),$$
where
$$
\opp{P}_{0}(p^{\mu}) = p^{-1}\slashed{p},
$$
hence and $p = \sqrt{p_{\mu}p^{\mu}}$.  Also $\dual{\psi} =\bar{\psi}$ and so $\dual{P}_0(p^{\mu}) = p^{-1 \dagger} \slashed{p}$. We take $\sqrt{p^{\mu}p_{\mu}} = i\vert p \vert$.  Thus if $p_{\mu}p^{\mu} > 0$:
$$
\opp{P}(p^{\mu}) = p^{-1}\slashed{p},
$$
and if $p_{\mu}p^{\mu} < 0$:
$$
\opp{P}(p^{\mu}) = \gamma^{5} \vert p \vert^{-1} \slashed{p}.
$$
The operator $P(p^{\mu})$ and hence the propagator, $G_{\rm F}(p;m)$ is well-defined and non-singular for all $p^2 \neq 0$ and $p^2 \neq m^2$. Singularities are only problematic if they result in singular contributions to the Green's function $G_{\rm F}(x-y;m)$.  We must show that $G_{\rm F}(x-y)$ is well-defined.  We have already noted that, with an appropriate choice of integration contour, the $p = \pm m$ singularities in $G_{\rm F}(p;m)$ are harmless.  What remains is to consider the new singularity at $p^2=0$.  

Let us write $p^{\mu}=(\omega, \mathbf{p}^{i})^{T}$, and then $p = \sqrt{\omega^2 - \mathbf{p}^2}$ and suppose that $A(\omega,\mathbf{p})$ is non-singular (or perhaps simply non-singular at $p=0$). We now consider the integral of $p^{-1}A(\omega,\mathbf{p})$ over all $\omega$:
\be
I_{A} \equiv \int p^{-1}A(\omega,\mathbf{p}) \omega =  \int_{-\infty}^{\infty} \frac{\dd \omega}{\sqrt{(\omega-\vert\mathbf{p}\vert)}\sqrt{(\omega+\vert\mathbf{p}\vert)}} A(\omega,\mathbf{p}).  \nonumber
\ee
We note that the simple pole in $p$ we have introduced does not result in a simple pole in $\omega$ or $\vert \mathbf{p}\vert$. Instead, as one approaches the points $p=0 \rightarrow \omega = \pm \mathbf{p}$ the integrand diverges more slowly than a simple pole and so the integral does not diverge. It should be noted though that the full specification of $p = \sqrt{p^2}$ (and hence $\opp{P}(x^{\mu})$) requires a choice of branch its value when $p^2 < 0$. This amounts to fixing the sign of $\sqrt{-1} = \pm i$.  Since our $\opp{P}(p^{\mu}) = \pm p_{\pm}^{-1}\slashed{p}$ the full specification of this operator must include a such a choice of branch.  Once the action of $p^{-1}$ on $p$ such that $p^2 < 0$ has been fixed,  integrals of the form of $I_{A}$ are well-defined.

Now $p = \sqrt{\omega^2 - \mathbf{p}^2}$ so with an appropriate choice of branch which fixes the sign of $\sqrt{-1} = \pm i$, we have that $p$ takes values from $\pm i \vert \mathbf{p}\vert$ to $0$ along the imaginary axis and then from $0$ to $\infty$ along the real-axis.

We also define $E(p^2,\mathbf{p}^2) = \sqrt{p^2+\mathbf{p}^2}$, and note that for fixed $p$ and $\mathbf{p}$, $\omega = \pm E(p^2,\mathbf{p}^2)$.  We also have $p\dd p = \omega \dd \omega$ and so:
\be
I_{A} &=& \int_{-\infty}^{\infty} p^{-1}A(\omega;\mathbf{p}) \dd \omega \\ &=& \mp i\int_{0}^{\vert\mathbf{p}\vert} \dd s \left[\frac{A(E(-s^2;\mathbf{p}^2);\mathbf{p})}{{E(-s^2,\mathbf{p}^2)}}+\frac{A(-E(-s^2;\mathbf{p}^2);\mathbf{p})}{{E(-s^2,\mathbf{p}^2)}}\right]\nonumber \\
&&+ \int_{0}^{\infty} \dd p \left[\frac{A(E(p^2;\mathbf{p^2});\mathbf{p})}{{E(p^2,\mathbf{p}^2)}}+\frac{A(-E(p^2;\mathbf{p}^2);\mathbf{p})}{{E(p^2,\mathbf{p}^2)}}\right]. \nonumber
\ee
The integrand on the right hand side is now manifestly finite at $p=0$. It is now clear that the $p^{-1}$ term in the integrand on the left hand side does not introduce a divergence in $I_{A}$.

Hence, using $\slashed{p} = p_{\mu}\gamma^{\mu} = \omega \gamma^{0} - \mathbf{p}^{i}\gamma^{i}$ and defining $x^{\mu}-y^{\mu} = (\tau,\mathbf{z}^{i})$:
\be
G_{F}(x-y) &=&  \frac{1}{(2\pi)^4} \lim_{\epsilon \rightarrow 0}\int \dd^3 \mathbf{p} e^{-i \mathbf{p}\cdot \mathbf{z}} \left\lbrace   I_{-}(\mathbf{p}) + I_{+}(\mathbf{p})\right\rbrace,
\ee
where
\be
I_{-}(\mathbf{p})  &=&  \int_{0}^{\vert \mathbf{p}\vert} \frac{\dd s}{-s^2-m^2+i\epsilon} \, \left[\mp i\gamma^{5}\gamma^{0}\sin(E(-s^2,\vert \mathbf{p}\vert^2) \tau)\right.\nonumber \\ &&+ \left.\left( \frac{s \mathbb{I}}{E(-s^2,\vert \mathbf{p}\vert^2)} \mp \frac{\gamma^{5}\mathbf{p}^{i}\gamma^{i}}{E(-s^2,\vert \mathbf{p}\vert^2)} \right)\cos(E(-s^2,\vert \mathbf{p}\vert^2) \tau)  \right], \nonumber \\
I_{+}(\mathbf{p})  &=&  \int_{0}^{\infty} \frac{\dd p}{p^2-m^2+i\epsilon} \, \left[-i\gamma^{0}\sin(E(p^2,\vert \mathbf{p}\vert^2) \tau)\right.\nonumber \\ &&+ \left.\left( \frac{p \mathbb{I}}{E(p^2,\vert \mathbf{p}\vert^2)} + \frac{ \mathbf{p}^{i}\gamma^{i}}{E(p^2,\vert \mathbf{p}\vert^2)} \right)\cos(E(p^2,\vert \mathbf{p}\vert^2) \tau)  \right]. \nonumber
\ee
At $p^2=0$, $E(p^2,\mathbf{p}^2) = \vert \mathbf{p}\vert$.  It is therefore clear that  the integrands of both $I_{-}$ and $I_{+}$ are non-singular as $p \rightarrow 0$ for all $\mathbf{p}$.  It follows that, despite the singular behaviour of $G_{F}(p;m)$ at $p=0$, the position space Green's function $G_{F}(x-y)$ is well-defined. This is enough to ensure that the pole at $p=0$ in $G_{F}(p;m)$ does not introduce any new divergent behaviour into quantum expectations and hence observables. Once the operator $P(x^{\mu})$ has been completely specified, which includes specifying the branch, the quantum theory is well-defined. 

This remains true when perturbatively renormalizable interactions are introduced, and in \S \ref{app:quant:int} below we show explicitly that the $p=0$ pole does not introduce any new divergences into loop integrals.  

\subsection{Quantum Effective Action}
For completeness, we finish our treatment of the free-field theory by considering the quantum effective action.  In flat-space, we define the energy functional:
\be
\mathcal{E}[J,\dual{J}\,] = i\ln Z[J,\dual{J}\,] = \iint \dual{J}(x) G_{F}(x-y)J(y)\dd^4 x\,\dd^4 y.
\ee
and define:
\be
\psi_{J}(x) &=& -i\frac{\delta \ln Z}{\delta \dual{J}} = -\frac{\delta \mathcal{E}}{\delta \dual{J}} = -\int \dd^4 y\, G_{F}(x-y)J(y) , \\
\dual{\psi}_{J}(x) &=& -i\frac{\delta \ln Z}{\delta J} = -\frac{\delta \mathcal{E}}{\delta J} = -\int \dd^4 y\, \dual{J}(y)G_{F}(y-x).
\ee
The quantum effective action, $\Gamma$, is defined thus:
\be
\Gamma[\psi_{J},\dual{\psi}_{J}] &=& -\int \dd^4 x\, \left[\dual{J}(x) \psi_{J}(x)+\dual{\psi}_{J}(x)J(x)\right]-\mathcal{E}[J,\dual{J}] \\ &=& \iint \dd^4 x\,\dd^4 y\, \dual{J}(y) G_{F}(x-y)J(x). \nonumber
\ee
Using the following equations:
\be
\left[-\slashed{\nabla}^2-m^2\right]\psi_{J}(x) = -\opp{P}_{+}(x) J(x),\nonumber \\
\bar{\psi}_{J}(x)\left[-\overleftarrow{\slashed{\nabla}}^2-m^2\right] = -\dual{J}(x)  \oppL{P}_{+}(x), \nonumber
\ee
we find:
\be
\Gamma[\psi,\dual{\psi}] = \iint \dual{\psi}(x)\oppL{P}_{+}(x)\left[-\slashed{\nabla}^2 -m^2\right]\opp{P}_{+}(x)\psi(x)\, \dd^4 x. \nonumber
\ee
Thus, as should be expected, the quantum effective action for the free field theory is equivalent to that of the classical theory (once the auxiliary $\chi$ and $\dual{\chi}$ fields have been integrated out).

\subsection{Interaction Terms}\label{app:quant:int}
Interaction terms can then be introduced as a perturbation about the free field theory by expanding out the exponential of $I_{\psi}$ in Eq. (\ref{Zeqn}) in the usual fashion. In flat-space this expansion is given by:
\be
Z[J,\dual{J}\,] = \left[1-i I_{\psi}\left(\frac{\delta}{i\delta J},\frac{\delta}{i\delta \dual{J}}\right) + I^2_{\psi}\left(\frac{\delta}{i\delta J},\frac{\delta}{i\delta \dual{J}}\right) + ...\right]Z_0[J,\dual{J}\,]. \label{Zexp}
\ee
From this one can extract Feymann rules in the usual manner and calculate $Z[\dual{J},\dual{J}]$.  The quantum theory derived from this procedure will be well-defined provided the form of the interactions is perturbatively renormalizable and provided the free field Green's function $G_{F}(x-y)$ is well-defined.  

We noted above that in our Lorentz invariant non-standard spinor model, the Fourier transform, $G_{\rm F}(p;m)$, of $G_{F}(x-y)$, had an additional simple pole at $p=0$ coming from $\opp{P}_0(p^{\mu}) = p^{-1}\slashed{p}$ terms in $\opp{P}$.  Crucially, however, we showed that this pole did not result in a divergent $G_{\rm F}(x-y)$; the Green's function remained well-defined.   It follows  Eq. (\ref{Zexp}) that, provided the interactions are perturbatively renormalizable, a non-free Lorentz invariant non-standard spinor will also have a well-defined quantum theory.

We now show this explicitly by noting that the $p=0$ pole in $G_{\rm F}(p;m)$ does not introduce any additional divergences into quantum loop integrals, and hence the pole in $\opp{P}(p^{\mu})$ is harmless.

Consider the contribution, $I_{\rm loop}$, to a general quantum loop integral from the integration over one of the loop momenta $p^{\mu}$. We define the number of internal lines in the Feymann diagram corresponding to this integral to be $N$.  Each internal line is represented in the integral by a propagator,  $G_{\rm F}(q_{i};m)$ where $q_{i}^{\mu} = p^{\mu}+k_{i}^{\mu}$ where we define $p^{\mu}$ so that $k_{0}^{\mu} = 0$ and the other $k_{i}^{\mu}$ are independent of each other.  The internal lines join at vertices which are represents by some matrix valued operators $\mathcal{M}_{(i)}{}^{AB\dots...}_{CD \dots...}(q_{j})$.  The capital letter indices may represent either space-time or spinor valued indices.  Henceforth we suppress these indices.

Thus we have:
\be
I_{\rm loop} = \int \frac{\dd^4 p}{(2\pi)^4}  \prod_{i=0}^{N-1} \left(\mathcal{M}_{i} G_{\rm F}(q^{\mu}_{i};m)\right) \mathcal{M}_{N}.
\ee
Now $\dd^4 p = \dd^4 q_{i}$ for any of the $q_{i}$ and so defining $q_{i}^{\mu} = (\omega_{i},\mathbf{q}_{i})$ and letting $E(q_{i}^2,\vert\mathbf{q}_{i}\vert^2) = \sqrt{q^2_{i}+\vert \mathbf{q}_{i}\vert^2}$ we have:
\be
I_{\rm loop} &=& \int \frac{\dd q_{j} \dd^3 \mathbf{q}_{j}}{E(q_{j},\mathbf{q}_{j})} \left\lbrace \left.X(q_{i};j)\right\vert_{\omega_{j} =E(q_{j},\mathbf{q}_{j})} +  \left.X(q_{i};j)\right\vert_{\omega_{j} =-E(q_{j},\mathbf{q}_{j})}\right \rbrace, \\
X(q_{i};j)  &=& \left[\prod_{i=0}^{j-1}\left(\mathcal{M}_{i} G_{F}(q_{i}^{\mu};m)\right)\right]\mathcal{M}_{j} \left[q_{j}G_{F}(q_{j}^{\mu};m)\right]\left[\prod_{k=j+1}^{N-1}\left(\mathcal{M}_{k} G_{F}(q_{k}^{\mu};m)\right)\right] \mathcal{M}_{N}. \nonumber
\ee
Note that in the integral, the integration along $q_{j}$ is along a path, $\gamma_{q}$, that runs along the imaginary axis from $\sqrt{-\vert \mathbf{q}_{j}\vert}$ to $0$ and then from $0$ to $\infty$ along the real axis.  We note that, for any $j$, $\left[q_{j}G_{F}(q_{j}^{\mu};m)\right]$ is finite as $q_{j}\rightarrow 0$, and  $\lim_{q_{j}\rightarrow 0} X(q_{i};j)$ is also finite for any $j$.  Therefore, for all $j$, the integrand of $I_{\rm loop}$ does not diverge near $q_{j}=0$ for all $j$.  The additional pole in $p$ the propagator does not lead to any new divergences in loop diagrams and hence in quantum expectations and observables.

\end{document}